\UseRawInputEncoding
\documentclass[prl, twocolumn, superscriptaddress, longbibliography]{revtex4-1}

\usepackage{dsfont}
\usepackage{amsmath}
\usepackage{epsfig}
\usepackage{graphicx}
\usepackage{bm}
\usepackage{amssymb}
\usepackage{slashed}
\usepackage{hyperref}
\hypersetup{
     colorlinks   = true,
     citecolor    = blue,
     urlcolor = blue,
    linkcolor = blue
}

\newcommand{\bnabla}{{\boldsymbol \nabla}}

\begin{document}
% ==============================================================================

\title{Preformed Cooper pairs in flat-band semimetals}

\author{Alexander~A.~Zyuzin}
\affiliation{Department of Applied Physics, Aalto University, FI-00076 AALTO, Finland}

\author{A.~Yu.~Zyuzin}
\affiliation{Ioffe Physical--Technical Institute,~194021 St.~Petersburg, Russia}

% ==============================================================================
\begin{abstract}
We study conditions for the emergence of the preformed Cooper pairs in materials hosting flat bands. As a particular example, we consider a semimetal, 
with a pair of three-band crossing points at which a flat band intersects with a Dirac cone, and focus on the $s$-wave intervalley pairing channel. 
The nearly dispersionless nature of the flat band at strong attraction between electrons promotes local Cooper pair formation so that the system may be modeled as an array of superconducting grains. 
Due to dispersive bands, Andreev scattering between the grains gives rise to the global phase-coherent superconductivity at low temperatures. We develop a mean-field theory to calculate transition temperature 
between the preformed Cooper pair state and the phase-coherent state for different interaction strengths in the Cooper channel. The transition temperature between semimetal and preformed Cooper pair phases is proportional
to the interaction constant, the dependence of the transition temperature to the phase-coherent state on the interaction constant is weaker. 
\end{abstract}
\maketitle

In condensed matter systems the nearly dispersionless flat band electronic structure may stimulate the interaction-induced instabilities.
Of particular interest is the interplay between flat band and superconductivity. The reason for that is the relatively large value of the superconducting transition temperature, which can be linearly proportional to the pairing interaction strength as was proposed by Khodel' and Shaginyan \cite{Khodel_Shaginyan} and later studied for instance in Refs. \cite{PhysRevLett_Masanori, Physica_C_Furukawa, PhysRevB_Kopnin_Heikkila_Volovik, PhysRevB_Nandkishore, Heikkila_twisted, Wu_Macdonald_twisted}. 

The examples of flat-band systems include multilayer graphene with rhombohedral stacking \cite{PhysRevB_Kopnin_Heikkila_Volovik}, interfaces between the domains in graphene with Bernal stacking order \cite{Bernal_graphite_Volovik}, twisted bilayer graphene \cite{PhysRevLetT_castro, PNAS_macdonald}, and semimetals with integer pseudospin quasiparticles \cite{PhysRevB_Dora, PhysRevB_Manes_3fold_theory, Science_Bernevig_Multifold_theory}. The latter is characterized by the existence of multiple-band crossing points at which the flat band intersects with the Dirac cones. For example, the low-energy electron excitations can be described by the Hamiltonian for a pseudospin-one particle, see Ref. \cite{Science_Bernevig_Multifold_theory}. 

Recently superconductivity has been observed in twisted bilayer graphene \cite{Exp_TBG_superconductor} and in Bernal bilayer graphene subject to applied perpendicular electric field \cite{AB_stacking_SC}. Signatures of superconductivity have been observed in highly oriented pyrolytic graphite \cite{Bernal_graphite_Volovik, Volovik_flat_band_perspective}.
Although, the semimetals hosting three-band touching points (and among them CoSi and RhSi) have been discovered \cite{PhysRevLett_Takane_Multifold_exp, Nature_Zhicheng_Multifold_exp, Nature_Sanchez_Multifold_exp} (Ref. \cite{RevModPhys_Ding} for a review) and several flat-band enhanced Cooper pairing channels have been explored theoretically \cite{PhysRevB_Nandkishore, PhysRevResearch_Lin}, superconductivity has not been detected yet.
Despite intensive research, the role of flat band in the Cooper pairing is far from being understood \cite{Volovik_flat_band_perspective}.

We emphasize that the effect of the flat band on the formation of superconductivity can be twofold. On one hand, the strong enhancement of the electronic density of states leads to higher critical temperatures of Cooper pairing. 
On the other, its nearly dispersionless nature can be a serious impediment to pair condensation. The flat band favors the localization of quasiparticles, which suppresses the superconducting phase stiffness.
It works against the long-range coherence leading rather to a situation with preformed Cooper pairing \cite{Nature_Sacepe_Feigelman}.

The problem of flat-band induced correlations between the Cooper pairs was analysed in Ref. \cite{Peotta_Torma}. The flat-band contribution to superconducting phase stiffness was shown to be finite and
originate from the position-dependent matrix structure of the respective wave function.
It is now believed that this contribution might eventually support the pair condensation. 
However, we argue that Ref. \cite{Peotta_Torma} overlooks superconductivity and deals with the preformed Cooper pair phase and the 
properties of the local pairs. In this theory, the flat-band contribution to phase stiffness results in narrow-range spatial correlations on the scale of the size of the preformed Cooper pair itself.
Instead, we expect different situation, in which local Cooper pairs coexist with the Fermi liquid. 

In our context, however, the Cooper pair formation and their condensation occur at different temperatures \cite{JLTP_Schmitt, Zyuzin_preformed}. 
In contrast to the previous research \cite{Khodel_Shaginyan, PhysRevLett_Masanori, Physica_C_Furukawa, PhysRevB_Kopnin_Heikkila_Volovik, PhysRevB_Nandkishore, Heikkila_twisted, Wu_Macdonald_twisted}, we emphasize the importance of both localized and delocalized quasiparticles on the emergence of superconductivity.
In addition to the flat band, materials inevitably host dispersive bands as well, which essentially contribute to the pair condensation.
Such situation exists in considered three-band semimetal.
We note that our theory might be extended to explain superconductivity in bilayer graphene with twisted and Bernal stacking.

We show that with the increase of electron-electron attraction, the system reaches a state, which can be modeled by an emergent granularity. 
It can be described by the Cooper pairs localized inside the grains lacking the long-range coherence. 
The superconducting order parameter exhibits strong spatial fluctuations.
The long-ranged Andreev coupling between the grains, thanks to the contribution of dispersive bands, establishes a coherent state at a lower temperature. 
We develop a mean-field theory to calculate the temperatures of Cooper pairs formation and their consecutive condensation.

\begin{figure}[t!]
\centering
\includegraphics[width=6.0cm]{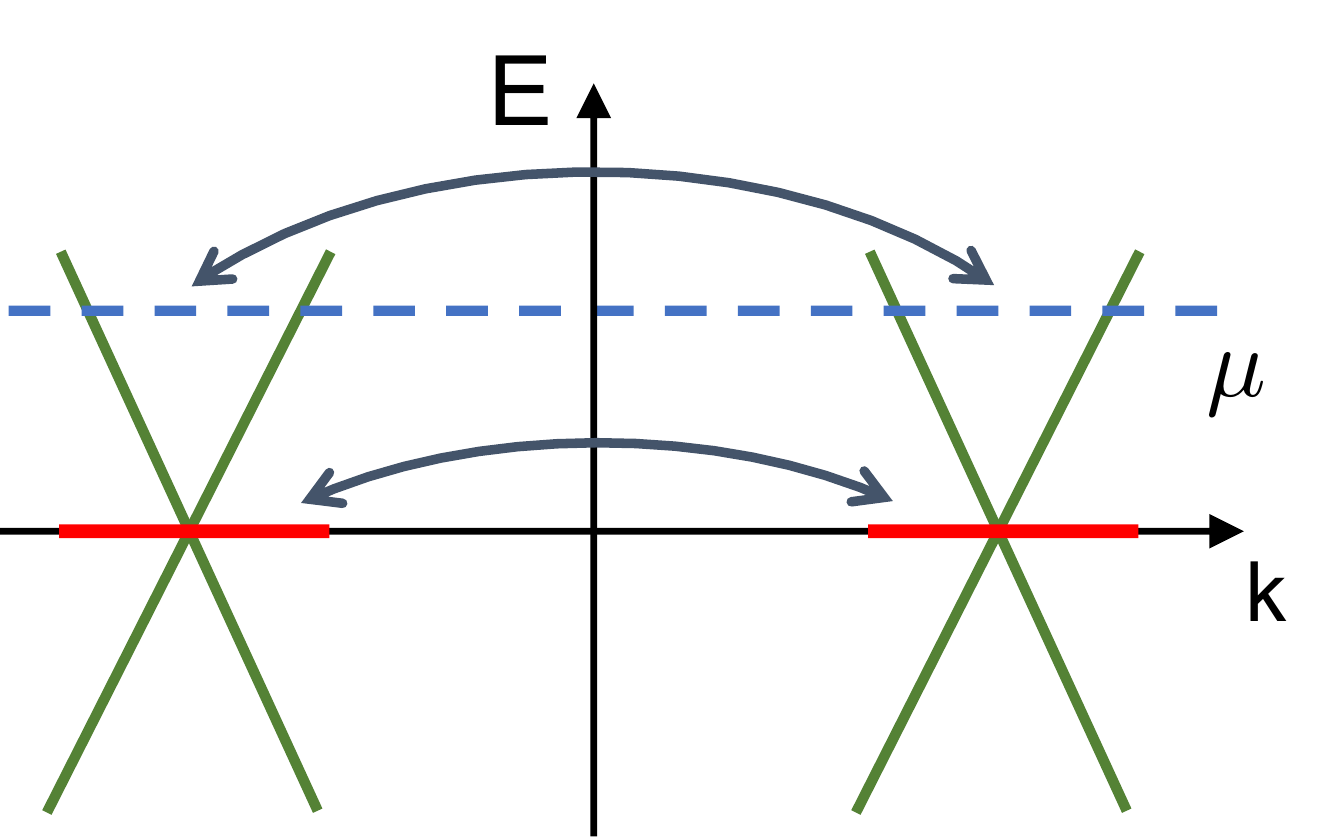}
\caption{\label{fig1} Schematics of the band structure $E(\mathbf{k})$ in the vicinity of two three-band-touching points at chemical potential $\mu$. 
There are two points at which Dirac cones and flat bands intersect. Superconducting pairing of electrons from different valleys is considered.}
\end{figure}

% ==============
\emph{Model of semimetal.}
% ==============
We consider a time-reversal symmetric semimetal with a pair of three-band crossing points at momenta $\pm \mathbf{K}_{\mathrm{D}}$ in the first Brillouin zone as shown schematically in Fig. (\ref{fig1}). As we ignore the single-particle intervalley scattering processes, the model Hamiltonian can be represented via a sum of two independent contributions from two valleys \cite{Science_Bernevig_Multifold_theory}: $\mathcal{H} = \int_{\mathbf{k}} \sum_{s=\pm}\Psi^{\dag}_{s,\mathbf{k}} v_{\mathrm{F}}\mathbf{S}\cdot \mathbf{k} \Psi_{s,\mathbf{k}}$, where $v_{\mathrm{F}}$ is the Fermi velocity, $\mathbf{k}$ is the momentum measured relatively to the $\pm \mathbf{K}_{\mathrm{D}}$ with $k \ll K_{\mathrm{D}}$, $\int_{\bf k}(..) \equiv \int \frac{d\mathbf{k}}{(2\pi)^3}(..)$, and $\mathbf{S} = (S_x,S_y,S_z)$ are the Gell-Mann matrices acting on "which band" pseudospin degree of freedom; see the Supplemental Material \cite{SM_editors}.
The electron operators are defined by $\Psi_{s,\mathbf{k}} = [\Psi_{s, +1,\mathbf{k}}, \Psi_{s, 0,\mathbf{k}}, \Psi_{s, -1,\mathbf{k}}]^{T}$, where indices $\pm 1, 0$ correspond to three different bands,
two of which are dispersive, $E_{\pm 1} = \pm v_{\mathrm{F}} k$, and another is flat, $E_{ 0} = 0$. The latter is considered in 
the infinite mass limit approximation, so that higher order momentum corrections are neglected. We will be using $\hbar = k_{\mathrm{B}} = 1$ units throughout the paper.

To analyze superconducting instability in the system, we introduce electron Green function in Matsubara representation  $G(\mathbf{r}, i\omega)  = \int_{\bf k}  G(\mathbf{k}, i\omega) e^{i\mathbf{k}\cdot\mathbf{r}}$, where $\omega=(2n + 1)\pi T $ is the Matsubara frequency at temperature $T$. The Green's function $G(\mathbf{k}, i\omega) = (i\omega -v_{\mathrm{F}}\mathbf{S}\cdot\mathbf{k} + \mu)^{-1}$ can be expressed as \cite{SM_editors}
\begin{eqnarray}\label{Green_Function}
G(\mathbf{k}, i\omega) = \frac{1- (\mathbf{S}\mathbf{n}_k)^2}{i\omega + \mu} + \frac{1}{2}\sum_{s=\pm1} \frac{ (\mathbf{S}\mathbf{n}_k)^2 + s(\mathbf{S}\mathbf{n}_k)}{i\omega + \mu - sv_{\mathrm{F}} k},~~~ 
\end{eqnarray}
where $\mu$ is the chemical potential and $\mathbf{n}_k = \mathbf{k}/k$ is a unit vector in the direction of momentum.
The chemical potential can be positive or negative, although we choose it to be positive since it does not change our result.
Attention shall be paid to the case of finite flat-band dispersion corrections, which violate the particle-hole symmetry \cite{PhysRevB_Nandkishore}. We will comment on that later in the conclusions. 

The first and second terms in (\ref{Green_Function}) describe contributions of the flat and dispersive bands, which can be separated into the local and 
nonlocal terms as $G(\mathbf{r}, i\omega) = G_{\mathrm{loc}}(\mathbf{r}, i\omega) + G_{\mathrm{nl}}(\mathbf{r}, i\omega)$, respectively. The local contribution is given by 
\begin{eqnarray}\label{local_GF}
G_{\mathrm{loc}}(\mathbf{r}, i\omega) = \frac{1}{i\omega + \mu} \left\{\delta(\mathbf{r}) + \frac{1}{4\pi r^3}\left[ 3(\mathbf{S}\mathbf{n}_r)^2 - \mathbf{S}^2\right] \right\},~~~
\end{eqnarray}
where now $\mathbf{n}_r = \mathbf{r}/r$ is the unit vector in coordinate space and $\delta(\mathbf{r})$ is the Dirac delta function in three dimension. 
The second dipole-like term decays as a cube of distance smearing the delta function. In the limit of $r\rightarrow 0$ the Green's function is cut by the interatomic distance.
We also note that the spatial and frequency dependent parts are separated in the flat-band model in the infinite mass approximation. 

It suffices to consider the nonlocal term in the limiting case, where $\mu \gg |\omega|$ and $\mu r/v_{\mathrm{F}} \gg 1$,
\begin{eqnarray}\label{nonlocal_GF}
G_{\mathrm{nl}}(\mathbf{r}, i\omega) = -\frac{\mu (\mathbf{S}\mathbf{n}_r) }{4\pi v_{\mathrm{F}}^2 r}[\mathrm{sgn}\omega + (\mathbf{S}\mathbf{n}_r) ]e^{-\frac{r}{v_{\mathrm{F}}}(\omega-i\mu) \mathrm{sgn}\omega}.~~~
\end{eqnarray}
The expected three-dimensional spatial coordinate dependence is supplemented by the unusual matrix structure. 
Let us now discuss superconductivity in flat band semimetal.

% ==============
\emph{Model of superconductivity.}
% ==============
We consider $s$-wave superconducting instability in the flat-band semimetal taking three-band semimetal as a particular example \cite{PhysRevB_Nandkishore}.  However, we note that our results are generally valid for systems with
coexisting dispersive and nearly flat bands. 
The symmetry analysis of the superconducting channels in three-band semimetal was performed in Refs.~\cite{PhysRevB_Nandkishore, PhysRevResearch_Lin}. Specifically, for clean systems possessing time-reversal symmetry, 
it was found that the flat band enhances intervalley Cooper pairing with total pseudospin $S=0$. The intervalley contribution to the interaction between electrons is given by \cite{SM_editors}
%\begin{eqnarray}\label{interaction_1}
%U = - \lambda \sum_{\{\alpha\}}\int_{\bf k, k'} \Psi_{1,\alpha_1, \mathbf{k}}^\dag  \Psi_{-1,\alpha_2, -\mathbf{k}}^\dag\Psi_{-1, \alpha_3, -\mathbf{k}'} \Psi_{1,\alpha_4, \mathbf{k}'},~~~~
%\end{eqnarray}
\begin{eqnarray}\label{interaction_1}
U = - \lambda \sum_{\alpha,\beta}\int_{\bf k, k'} (\Psi_{1,\alpha, \mathbf{k}}^\dag  \Psi_{1,\alpha, \mathbf{k}'}) ( \Psi_{-1,\beta, -\mathbf{k}}^\dag\Psi_{-1, \beta, -\mathbf{k}'}),~~~~
\end{eqnarray}
where $ \lambda > 0$ is the interaction constant. We seek for the case in which flat-band significantly contributes to superconductivity. Among many possible superconducting states we focus on the $s$-wave inter-valley 
odd pairing \cite{PhysRevB_Nandkishore, PhysRevResearch_Lin}.
The pairing channels can be distinguished by the total pseudospin $S$ of Cooper pairs. 
In our case, one can only have even $S =0$ and $S=2$ due to the Pauli principle. We focus on the $S =0$ channel, which has the highest superconducting transition temperature.
The extended analysis of superconducting states for $S=2$ in the model, which takes into account quadratic momentum corrections to the single-particle Hamiltonian, can be found in \cite{Herbut}.

%%%%%%%%%%%%%%%%%%%%%% BEGIN FIGURE %%%%%%%%%%%%%%%%%%%
%%%%%%%%%%%%%%%%%%%%%% END FIGURE %%%%%%%%%%%%%%%%%%%

Let us now qualitatively estimate the superconducting vertex part describing Cooper instability, 
\begin{equation}\label{Vertex_part}
\mathrm{det}[1-\lambda \Pi(\mathbf{q})]=0, 
\end{equation}
where
$
\Pi(\mathbf{q}) = T \sum_{\omega}\int_{\bf k} G(\mathbf{k}, i\omega) [G(\mathbf{q}-\mathbf{k}, -i\omega)|_{\mathbf{S}\rightarrow -\mathbf{S}}]
$ and summation is performed over the Matsubara frequencies.
Due to the local term in Green's function (\ref{local_GF}), the integrand in $\Pi(\mathbf{q})$ diverges at large wave-vectors. 
And it is convenient to single out non-local contributions (\ref{nonlocal_GF}), which contain the usual logarithmic ultraviolet cutoff. 
All in all, we separate local and non-local contributions $\Pi(\mathbf{q}) =\Pi_{\mathrm{loc}}(\mathbf{q}) + \Pi_{\mathrm{nl}}(\mathbf{q})$ and neglect crossed terms between them (as we are interested in the two limiting cases only).

Consider momentum expansion of the vertex part $\Pi(\mathbf{q}) \approx \Pi + q^2 \delta \Pi$, where the second term describes superconducting stiffness. The contribution of the local term in Green's function (\ref{local_GF}) to the superconducting vertex part $\Pi_{\mathrm{loc}} \sim (K^3/\mu)\mathrm{th}(\mu/2T)$ is proportional to 
the volume of the flat band in momentum space $K^3 \equiv \int_{\bf k}$.
Using (\ref{nonlocal_GF}) and assuming $\mu \gg |\omega|$, a straightforward calculation results in the nonlocal term $\Pi_{\mathrm{nl}} \sim \mu^2 \ln(\mu/T)/v_{\mathrm{F}}^3$. 
Let us compare two terms at $\mu>T$,
\begin{eqnarray}\label{Ratio}
\frac{\Pi_{\mathrm{loc}}}{\Pi_{\mathrm{nl}}} \sim \left(\frac{K}{p_{\mathrm{F}}}\right)^3 \sim \frac{1}{V_{\mathrm{G}}p_{\mathrm{F}}^{3}},~~
\end{eqnarray}
where we introduced an effective volume $V_{\mathrm{G}} \sim K^{-3}$.
Let us compare the momentum dependent corrections to the vertex parts as well. Taking $\delta\Pi_{\mathrm{loc}} \sim (K/\mu)\mathrm{th}(\mu/2T) $ and $\delta\Pi_{\mathrm{nl}} \sim \mu^2/(T^2v_{\mathrm{F}})$, we estimate
\begin{equation}\label{Stiffness}
\frac{\delta\Pi_{\mathrm{loc}}}{\delta\Pi_{\mathrm{nl}}} \sim \left(\frac{T}{\mu}\right)^2 (V_{\mathrm{G}}p_{\mathrm{F}}^{3})^{-1/3}.
\end{equation}
At $K \gg p_{\mathrm{F}}$, we may adopt a model of a granular system, in which each grain hosts a Cooper pair.
The typical volume of the grain is of the order of $V_{\mathrm{G}}$.

\begin{figure}[t!]
 \centering
\includegraphics[width=6cm]{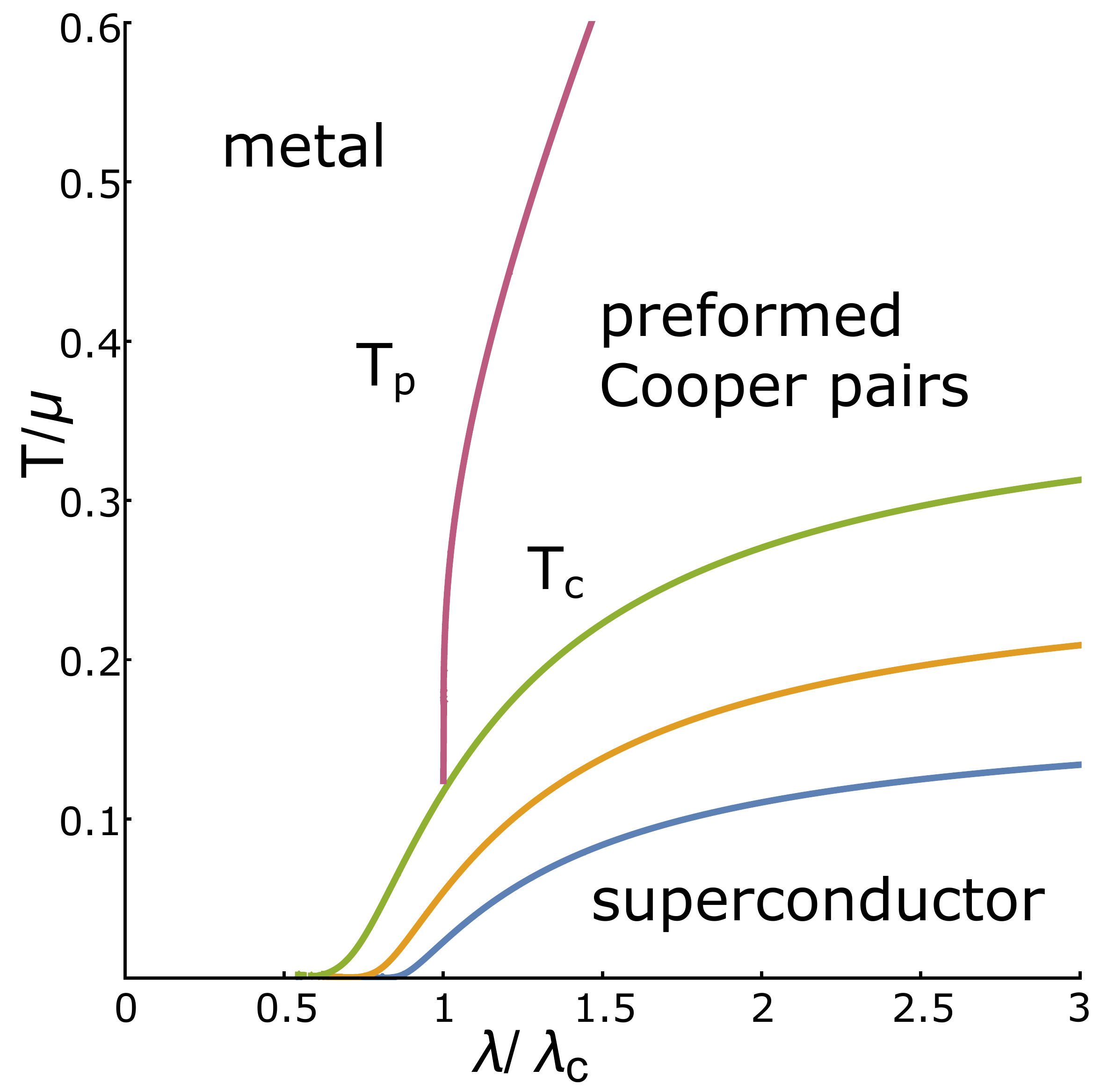}
\caption{\label{fig2} The phase diagram of doped semimetal, preformed Cooper pair, and phase-coherent states as a function of the interaction constant $\lambda$ (normalized by the critical value $\lambda_{\mathrm{c}}$) and temperature $T$ (normalized by the chemical potential $\mu$). The transition temperature $T_{\mathrm{p}} $ is linearly proportional to the interaction constant at $\lambda/\lambda_{\mathrm{c}} \gtrsim 1$ . Low-temperature curves describe the boundary of the phase-coherent state. Here the increase of dimensionless parameter $\lambda_{\mathrm{c}}\nu_{\mathrm{nl}}/3 = (0.05, 0.1, 0.2)$ increases $T_{\mathrm{c}}$. The tricritical point is shown for a single parameter only. }
\end{figure}

In this limit, at large interaction constant the local contribution $\Pi_{\mathrm{loc}}$ determines the instability towards the Cooper pair formation.
The dispersionless nature of the flat band prevents establishing global coherence in the system.
It rather leads to phase fluctuations of the order parameter on the scale of the size of the grain.
Although, by lowering the temperature, see Fig. (\ref{fig2}), one may reach a situation, in which the global coherence is fulfilled by long-range coupling between the grains.

% ==============
\emph{Ginzburg-Landau functional.}
% ==============
To proceed, we will analyze the superconducting instability within the Ginzburg-Landau (GL) functional framework in the static approximation \cite{Zyuzin_preformed}. We assume that the semimetal can be fragmented into a matrix of grains with equal volumes $V_{\mathrm{G}}$ and consider the situation in which the phase of the order parameter $\Delta_{i}$ (where index $i$ labels the grain) varies from grain to grain, while its amplitude is grain independent. In this model, the system is described by the Bogoliubov-deGennes (BdG) Hamiltonian 
$\mathcal{H} = \sum_{i}\int _{V_{\mathrm{G}}}d\mathbf{r} \Phi_i^\dag(\mathbf{r}) H_{i}(\mathbf{r})\Phi_i(\mathbf{r})$, where integration is performed over the volume of the grain $V_{\mathrm{G}}$ \cite{SM_editors},
\begin{equation}
H_{i}(\mathbf{r}) = \left[
 \begin{matrix}
-i v_{\mathrm{F}}\mathbf{S}\cdot \bnabla-\mu& \Delta_{i}(\mathbf{r})  \\
 \Delta_{i}^{*}(\mathbf{r}) & i v_{\mathrm{F}}\mathbf{S}\cdot \bnabla +\mu
 \end{matrix}
\right],
\end{equation}
and the Gorkov-Nambu operator on grain $i$ is given by $\Phi_i(\mathbf{r}) =  [\Psi^{T}_{1,i}(\mathbf{r}), \gamma\Psi^{*}_{-1,i}(\mathbf{r})]^T$.
Here a unitary operator $\gamma = e^{i\pi S_y}$ transforms the spin-1 operators as $\gamma^\dag \mathbf{S}^* \gamma = - \mathbf{S}$ \cite{PhysRevB_Nandkishore}. 
It resembles the antisymmetric property of the spin-matrix structure of the gap function in usual superconductors. Note that we neglect single-particle intervalley scattering processes, which results 
in the $6\times 6$ matrix structure of BdG Hamiltonian (similarly to the $2\times 2$ matrix structure reduction of the BdG Hamiltonian in usual superconductors).

In the limit of small gap function $|\Delta_i| \ll \mu$, the GL functional can be further expanded in powers of the order parameter. 
In this expansion, the superconducting phase stiffness consists of contributions from both local and nonlocal terms in the Green's function (\ref{Green_Function}).
Although noting (\ref{Stiffness}), the former is smaller compared to the nonlocal contribution, which allows us to neglect variation of the order parameter inside the grain and focus on the intergrain coupling only.

Taking into account both local (\ref{local_GF}) and nonlocal (\ref{nonlocal_GF}) contributions, the GL functional yields \cite{SM_editors}
\begin{eqnarray}\label{GL_functional_trunc}\nonumber
F &=& \sum_i F_{i}-\sum_{i\neq j} F_{i,j} \equiv V_{\mathrm{G}}\sum_i \bigg\{a |\Delta_i|^2 + \frac{b}{2}|\Delta_i|^4\bigg\}
\\
&-&\frac{\nu_{\mathrm{nl}}}{2v_{\mathrm{F}}}V_{\mathrm{G}}^2T\sum_{\omega}\sum_{i\neq j} \frac{e^{-\frac{2|\omega|}{v_{\mathrm{F}}}|\mathbf{r}_i-\mathbf{r}_j|}}{|\mathbf{r}_i - \mathbf{r}_j|^2} |\Delta_i\Delta_j |\cos(\phi_{ij}),~~~~
\end{eqnarray}
where $a = 3(\lambda^{-1} - \lambda_{\mathrm{c}}^{-1} \mathrm{th} \frac{\mu}{2T})$ and $b = 3\lambda_{\mathrm{c}}^{-1}(\mathrm{sh}\frac{\mu}{T}- \frac{\mu}{T})/(4\mu^2 \mathrm{ch}^2\frac{\mu}{2T})$ are the model dependent coefficients. We consider the case when the chemical potential $\mu$ is smaller than the Debye frequency.
It also suffices to introduce a critical value of interaction constant 
$
\lambda_{\mathrm{c}} = 6\mu/K^3
$.
The last term in (\ref{GL_functional_trunc}) describes long-range Andreev coupling between the grains, which is weighted by
the density of states per valley at the Fermi energy $\nu_{\mathrm{nl}} = \mu^2/(2\pi^2 v_{\mathrm{F}}^3)$. To obtain this term one follows familiar microscopic derivation within the GL formalism \cite{AGD}.
Note that the Andreev term is smaller than the second term in the coefficient $a$. The latter is defined by the flat-band contribution. We neglect weak corrections from delocalized states to the coefficient $a$ within the granular model.
Andreev coupling contributes to quartic terms in general form $\propto \Delta_i\Delta_j\Delta_k^*\Delta_{\ell}^*$, although these terms are small compared to $b|\Delta_i|^4$ in (\ref{GL_functional_trunc}).

Consider a situation in which weak Andreev coupling between the grains can be neglected. At $a<0$, from the extremum of (\ref{GL_functional_trunc}), we obtain nonzero local $|\Delta_i|$ with random phase.
We identify this case as preformed Cooper pair phase. In this case, equation $a=0$ determines the temperature of preformed Cooper pair formation on the grain.

Provided $\lambda \geq \lambda_{\mathrm{c}}$ one obtains
$
T_{\mathrm{p}} = \mu/2\mathrm{arcth}(\lambda_{\mathrm{c}}/\lambda)
$ \cite{PhysRevB_Nandkishore}.
This is the temperature of the phase transition between a doped semimetal and preformed Cooper pair state. The low-doping case $\mu \ll T$ requires large interaction constant $\lambda/\lambda_{\mathrm{c}} \gg 1$ for the transition. Here the critical temperature is proportional to the interaction constant and inversely proportional to the volume of preformed Cooper pair $T_{\mathrm{p}} = \lambda K^3/12$ \cite{Physica_C_Furukawa, PhysRevB_Kopnin_Heikkila_Volovik, PhysRevB_Nandkishore, Heikkila_twisted}. 

At high doping $\mu \gg T$, the transition takes place when the interaction constant is larger than the critical value, $\lambda \simeq \lambda_{\mathrm{c}}$, \cite{JLTP_Schmitt}.
In this limit the coefficients in (\ref{GL_functional_trunc}) can be simplified as $a = 3 (\lambda^{-1} - \lambda_{\mathrm{c}}^{-1})$ and $b=3/(2\mu^2\lambda_{\mathrm{c}})$. We shall focus on this case in what follows.
Let us now calculate the transition temperature to the phase-coherent state, which is driven by the Andreev coupling.

% ==============
\emph{Transition between preformed-pair and phase-coherent states.}
% ==============
With the increase of interaction constant $\lambda$, the impact of dispersive bands enhances Andreev coupling between the superconducting grains.
As a result, the system may reach the phase coherence. In what follows, we develop a mean-field theory to calculate the superconducting transition temperature.
%%%%%%%%%%%%%%%%%%%%%% BEGIN FIGURE %%%%%%%%%%%%%%%%%%%
%%%%%%%%%%%%%%%%%%%%%% END FIGURE %%%%%%%%%%%%%%%%%%%

Within the mean-field approach, the fluctuating values of the order parameter $\Delta_i$ are replaced by an average order parameter $\langle \Delta \rangle$ \cite{Zyuzin_preformed, SM_editors}.
The self-consistent mean-field equation is
\begin{eqnarray}\label{Self_consistent_eq}
\langle \Delta \rangle = \frac{\int \mathcal{D}\Delta \mathcal{D}\Delta^* \Delta_0 e^{- F/T}}{\int \mathcal{D}\Delta \mathcal{D}\Delta^* e^{- F/T}}\approx \frac{\int d\Delta d\Delta^* \Delta e^{- F_{MF}/T}}{\int d\Delta d\Delta^* e^{- F_{MF}/T}},~~~~~
\end{eqnarray}
in which $\mathcal{D}\Delta \equiv \Pi_{i}d\Delta_i$. The mean-field functional reads
\begin{eqnarray}\label{functional_MF}
F_{\mathrm{MF}} &=& F_{0}-\sum_{i\neq 0} F_{i,0}\\\nonumber
&=&
V_{\mathrm{G}}\left[a |\Delta|^2 +\frac{b}{2}|\Delta|^4 - c \left(\langle \Delta \rangle \Delta^* + \langle \Delta^* \rangle \Delta\right)\right].
\end{eqnarray}
In the continuum limit, we substitute $\sum_{i\neq 0}  = V_{\mathrm{G}}^{-1}\int_V d\mathbf{r}$ and obtain coefficient $c= \nu_{\mathrm{nl}}  \ln\left| \mu/T\right|$ within the logarithmic accuracy. 

Without loosing the generality, the averaged order parameter $\langle \Delta \rangle$ can be restricted to real value.
At $c\langle \Delta \rangle \ll \sqrt{|a|T/V_{\mathrm{G}}} \mathrm{max}(1, bT/a^2V_{\mathrm{G}})$, expanding integrands in Eq. (\ref{Self_consistent_eq}) 
in powers of $\langle \Delta \rangle$, we obtain
\begin{align}\label{Equation_Delta}\nonumber
&1 = \nu_{\mathrm{nl}}\frac{V_{\mathrm{G}}\langle |\Delta|^2 \rangle }{T}\ln\left|\frac{\mu}{T}\right|,\\
&\langle |\Delta|^2 \rangle = \frac{\int_0^{\infty} dx x e^{-\frac{V_{\mathrm{G}}}{T}(a x + \frac{b}{2}x^2 )}}{\int_0^{\infty} dx e^{- \frac{V_{\mathrm{G}}}{T}(a x + \frac{b}{2}x^2 )} }.
\end{align}
The solution to Eq. (\ref{Equation_Delta}) is shown in Fig. (\ref{fig2}). Analytical expressions can be analysed in several limiting cases.

First, consider a situation in which the interaction constant $\lambda$ is much smaller the critical value, $\lambda \ll \lambda_{\mathrm{c}}$, so that $\Delta_i =0$ ($a>0$).
In this weak coupling regime, the $b x^2$ term in formula (\ref{Equation_Delta}) can be neglected provided $a^2V_{\mathrm{G}}/b T \gg 1$. 
Performing integration in (\ref{Equation_Delta}) one obtains expression for the square of quasi-particle energy gap $\langle |\Delta|^2 \rangle \approx T/aV_{\mathrm{G}}$. 
As a result, the transition temperature to the coherent state is given by
\begin{equation}\label{Low}
T_{\mathrm{c}} = \mu \exp\left\{- \frac{3}{\lambda \nu_{\mathrm{nl}}}(1-\lambda/\lambda_{\mathrm{c}}) \right\}.
\end{equation}
This mean-field solution coincides with the exact BCS expression.
The flat band gives $ 1- \lambda/\lambda_{\mathrm{c}}$ enhancement correction in the exponent.

Second, consider a semimetal at the vicinity of the transition to preformed Cooper pair phase, $\lambda \approx \lambda_{\mathrm{c}}$. At $a^2V_{\mathrm{G}}/bT \ll 1$, we can neglect $a$-term compared with the nonlinear $b$-term in (\ref{Equation_Delta}) and obtain
$\langle |\Delta|^2 \rangle \approx \sqrt{2T/\pi b V_{\mathrm{G}} }$. 
Taking into account $V_{\mathrm{G}} K^3 \approx 1$ and using expressions for $\lambda_{\mathrm{c}}$ and $b$, we find
\begin{equation}
T_{\mathrm{c}} = \frac{2\mu}{9\pi} \left(\lambda_{\mathrm{c}}\nu_{\mathrm{nl}}\ln\left| \mu/T_{\mathrm{c}}\right| \right)^2.
\end{equation}
This result is valid for both signs of the coefficient $a$. Due to $\lambda_{\mathrm{c}}\nu_{\mathrm{nl}} < 1$, the $T_{c}$ is proportional to the critical value of the interaction constant squared.

Third, consider the preformed Cooper pair phase, $\lambda \gtrsim \lambda_{\mathrm{c}}$. For weak fluctuations $a^2V_{\mathrm{G}}/bT \gg 1$, using $\langle |\Delta|^2 \rangle \approx - a/b = 2\mu^2 (1-\lambda_{\mathrm{c}}/\lambda)$, we obtain 
\begin{equation}
T_{\mathrm{c}} = \frac{\lambda_{\mathrm{c}}\nu_{\mathrm{nl}}}{3}  \mu \left(1- \frac{\lambda_{\mathrm{c}}}{\lambda} \right)\ln\left| \frac{\mu}{T_{\mathrm{c}}}\right|.
\end{equation}
The transition temperature increases with the increase of $\lambda$. However, in the limit of $\lambda \gg \lambda_{\mathrm{c}}$, the GL expansion is no longer valid. The investigation of this case deserves a separate study.

% ==============
\emph{Conclusions.}
% ==============
Let us now briefly comment on the effect of finite $\propto k^2$ corrections to the Hamiltonian of semimetal. In this case, the flat band acquires a finite curvature.
As noted in Ref.~\cite{PhysRevB_Nandkishore} accounting for such term results in vanishing of the threshold value $\lambda_{\mathrm{c}}$, which is required for preformed Cooper pairing, provided the chemical potential crosses the band.
Hence, enhancement of the transition temperature $T_{\mathrm{c}}$ (\ref{Low}) at smaller values of the interaction constant $\lambda \rightarrow 0$ is expected for particular doping, which depends on the sign of $\propto k^2$ correction term.
We also note that materials may contain other dispersive bands, which can coexist with the Dirac cones at the chemical potential, and
contribute to the long-range coupling as well.

It would be interesting to extend the above-presented research to explain superconductivity in twisted bilayer graphene \cite{Exp_TBG_superconductor} and in graphite with Bernal stacking order \cite{Bernal_graphite_Volovik}. 
The moir\'e pattern can be modeled as a system of coupled grains \cite{Bernevig_flat_band}. We argue that in this situation the intergrain coupling leads to the phase-coherent state at temperatures lower than the temperature of the on-grain Cooper pair formation.
We will consider superconductivity in twisted bilayer graphene in future work.

To conclude, in this paper we have demonstrated that a nearly dispersionless flat band at strong attraction between electrons manifests itself in the emergent granularity and the Cooper pair pre-formation. 
The dispersive bands, which coexist with the flat bands, promote the global phase-coherent superconducting state at low temperatures. 
We have calculated the temperature of the phase transition between the preformed pairs and phase-coherent states in a semimetal hosting a pair of three-band crossing points.
Experimentally, the preformed Cooper pairs may be probed locally via low-temperature spectroscopy \cite{Nature_Sacepe_Feigelman}.

% ==============
%\textbf{Acknowledgements.}
% ==============
The authors are thankful to Vladimir Zyuzin for critical discussions and to Pirinem School of Theoretical Physics for warm hospitality.
This research was supported by the Academy of Finland (Project No. 308339) and in parts by the Academy of Finland Centre of Excellence program (Project No. 336810).

\bibliography{Flatband_references_arxiv.bib}

\begin{thebibliography}{29}%
\makeatletter
\providecommand \@ifxundefined [1]{%
 \@ifx{#1\undefined}
}%
\providecommand \@ifnum [1]{%
 \ifnum #1\expandafter \@firstoftwo
 \else \expandafter \@secondoftwo
 \fi
}%
\providecommand \@ifx [1]{%
 \ifx #1\expandafter \@firstoftwo
 \else \expandafter \@secondoftwo
 \fi
}%
\providecommand \natexlab [1]{#1}%
\providecommand \enquote  [1]{``#1''}%
\providecommand \bibnamefont  [1]{#1}%
\providecommand \bibfnamefont [1]{#1}%
\providecommand \citenamefont [1]{#1}%
\providecommand \href@noop [0]{\@secondoftwo}%
\providecommand \href [0]{\begingroup \@sanitize@url \@href}%
\providecommand \@href[1]{\@@startlink{#1}\@@href}%
\providecommand \@@href[1]{\endgroup#1\@@endlink}%
\providecommand \@sanitize@url [0]{\catcode `\\12\catcode `\$12\catcode
  `\&12\catcode `\#12\catcode `\^12\catcode `\_12\catcode `\%12\relax}%
\providecommand \@@startlink[1]{}%
\providecommand \@@endlink[0]{}%
\providecommand \url  [0]{\begingroup\@sanitize@url \@url }%
\providecommand \@url [1]{\endgroup\@href {#1}{\urlprefix }}%
\providecommand \urlprefix  [0]{URL }%
\providecommand \Eprint [0]{\href }%
\providecommand \doibase [0]{http://dx.doi.org/}%
\providecommand \selectlanguage [0]{\@gobble}%
\providecommand \bibinfo  [0]{\@secondoftwo}%
\providecommand \bibfield  [0]{\@secondoftwo}%
\providecommand \translation [1]{[#1]}%
\providecommand \BibitemOpen [0]{}%
\providecommand \bibitemStop [0]{}%
\providecommand \bibitemNoStop [0]{.\EOS\space}%
\providecommand \EOS [0]{\spacefactor3000\relax}%
\providecommand \BibitemShut  [1]{\csname bibitem#1\endcsname}%
\let\auto@bib@innerbib\@empty
%</preamble>
\bibitem [{\citenamefont {Khodel'}\ and\ \citenamefont
  {Shaginyan}(1990)}]{Khodel_Shaginyan}%
  \BibitemOpen
  \bibfield  {author} {\bibinfo {author} {\bibfnamefont {V.~A.}\ \bibnamefont
  {Khodel'}}\ and\ \bibinfo {author} {\bibfnamefont {V.~R.}\ \bibnamefont
  {Shaginyan}},\ }\bibfield  {title} {\enquote {\bibinfo {title}
  {{Superfluidity in system with fermion condensate}},}\ }\href
  {http://jetpletters.ru/ps/1143/article_17312.shtml} {\bibfield  {journal}
  {\bibinfo  {journal} {Jetp Lett.}\ }\textbf {\bibinfo {volume} {51}},\
  \bibinfo {pages} {553} (\bibinfo {year} {1990})}\BibitemShut {NoStop}%
\bibitem [{\citenamefont {Imada}\ and\ \citenamefont
  {Kohno}(2000)}]{PhysRevLett_Masanori}%
  \BibitemOpen
  \bibfield  {author} {\bibinfo {author} {\bibfnamefont {M.}~\bibnamefont
  {Imada}}\ and\ \bibinfo {author} {\bibfnamefont {M.}~\bibnamefont {Kohno}},\
  }\bibfield  {title} {\enquote {\bibinfo {title} {{Superconductivity from Flat
  Dispersion Designed in Doped Mott Insulators}},}\ }\href {\doibase
  10.1103/PhysRevLett.84.143} {\bibfield  {journal} {\bibinfo  {journal} {Phys.
  Rev. Lett.}\ }\textbf {\bibinfo {volume} {84}},\ \bibinfo {pages} {143--146}
  (\bibinfo {year} {2000})}\BibitemShut {NoStop}%
\bibitem [{\citenamefont {Miyahara}\ \emph {et~al.}(2007)\citenamefont
  {Miyahara}, \citenamefont {Kusuta},\ and\ \citenamefont
  {Furukawa}}]{Physica_C_Furukawa}%
  \BibitemOpen
  \bibfield  {author} {\bibinfo {author} {\bibfnamefont {S.}~\bibnamefont
  {Miyahara}}, \bibinfo {author} {\bibfnamefont {S.}~\bibnamefont {Kusuta}}, \
  and\ \bibinfo {author} {\bibfnamefont {N.}~\bibnamefont {Furukawa}},\
  }\bibfield  {title} {\enquote {\bibinfo {title} {{BCS theory on a flat band
  lattice}},}\ }\href {\doibase https://doi.org/10.1016/j.physc.2007.03.393}
  {\bibfield  {journal} {\bibinfo  {journal} {Physica C: Superconductivity}\
  }\textbf {\bibinfo {volume} {460-462}},\ \bibinfo {pages} {1145--1146}
  (\bibinfo {year} {2007})},\ \bibinfo {note} {proceedings of the 8th
  International Conference on Materials and Mechanisms of Superconductivity and
  High Temperature Superconductors}\BibitemShut {NoStop}%
\bibitem [{\citenamefont {Kopnin}\ \emph {et~al.}(2011)\citenamefont {Kopnin},
  \citenamefont {Heikkil\"a},\ and\ \citenamefont
  {Volovik}}]{PhysRevB_Kopnin_Heikkila_Volovik}%
  \BibitemOpen
  \bibfield  {author} {\bibinfo {author} {\bibfnamefont {N.~B.}\ \bibnamefont
  {Kopnin}}, \bibinfo {author} {\bibfnamefont {T.~T.}\ \bibnamefont
  {Heikkil\"a}}, \ and\ \bibinfo {author} {\bibfnamefont {G.~E.}\ \bibnamefont
  {Volovik}},\ }\bibfield  {title} {\enquote {\bibinfo {title}
  {{High-temperature surface superconductivity in topological flat-band
  systems}},}\ }\href {\doibase 10.1103/PhysRevB.83.220503} {\bibfield
  {journal} {\bibinfo  {journal} {Phys. Rev. B}\ }\textbf {\bibinfo {volume}
  {83}},\ \bibinfo {pages} {220503(R)} (\bibinfo {year} {2011})}\BibitemShut
  {NoStop}%
\bibitem [{\citenamefont {Lin}\ and\ \citenamefont
  {Nandkishore}(2018)}]{PhysRevB_Nandkishore}%
  \BibitemOpen
  \bibfield  {author} {\bibinfo {author} {\bibfnamefont {Yu-Ping}\ \bibnamefont
  {Lin}}\ and\ \bibinfo {author} {\bibfnamefont {R.~M.}\ \bibnamefont
  {Nandkishore}},\ }\bibfield  {title} {\enquote {\bibinfo {title} {{Exotic
  superconductivity with enhanced energy scales in materials with three band
  crossings}},}\ }\href {\doibase 10.1103/PhysRevB.97.134521} {\bibfield
  {journal} {\bibinfo  {journal} {Phys. Rev. B}\ }\textbf {\bibinfo {volume}
  {97}},\ \bibinfo {pages} {134521} (\bibinfo {year} {2018})}\BibitemShut
  {NoStop}%
\bibitem [{\citenamefont {Peltonen}\ \emph {et~al.}(2018)\citenamefont
  {Peltonen}, \citenamefont {Ojaj\"arvi},\ and\ \citenamefont
  {Heikkil\"a}}]{Heikkila_twisted}%
  \BibitemOpen
  \bibfield  {author} {\bibinfo {author} {\bibfnamefont {T.~J.}\ \bibnamefont
  {Peltonen}}, \bibinfo {author} {\bibfnamefont {R.}~\bibnamefont
  {Ojaj\"arvi}}, \ and\ \bibinfo {author} {\bibfnamefont {T.~T.}\ \bibnamefont
  {Heikkil\"a}},\ }\bibfield  {title} {\enquote {\bibinfo {title} {{Mean-field
  theory for superconductivity in twisted bilayer graphene}},}\ }\href
  {\doibase 10.1103/PhysRevB.98.220504} {\bibfield  {journal} {\bibinfo
  {journal} {Phys. Rev. B}\ }\textbf {\bibinfo {volume} {98}},\ \bibinfo
  {pages} {220504(R)} (\bibinfo {year} {2018})}\BibitemShut {NoStop}%
\bibitem [{\citenamefont {Wu}\ \emph {et~al.}(2018)\citenamefont {Wu},
  \citenamefont {MacDonald},\ and\ \citenamefont
  {Martin}}]{Wu_Macdonald_twisted}%
  \BibitemOpen
  \bibfield  {author} {\bibinfo {author} {\bibfnamefont {F.}~\bibnamefont
  {Wu}}, \bibinfo {author} {\bibfnamefont {A.~H.}\ \bibnamefont {MacDonald}}, \
  and\ \bibinfo {author} {\bibfnamefont {I.}~\bibnamefont {Martin}},\
  }\bibfield  {title} {\enquote {\bibinfo {title} {{Theory of Phonon-Mediated
  Superconductivity in Twisted Bilayer Graphene}},}\ }\href {\doibase
  10.1103/PhysRevLett.121.257001} {\bibfield  {journal} {\bibinfo  {journal}
  {Phys. Rev. Lett.}\ }\textbf {\bibinfo {volume} {121}},\ \bibinfo {pages}
  {257001} (\bibinfo {year} {2018})}\BibitemShut {NoStop}%
\bibitem [{\citenamefont {Esquinazi}\ \emph {et~al.}(2014)\citenamefont
  {Esquinazi}, \citenamefont {Heikkil\"a}, \citenamefont {Lysogorskiy},
  \citenamefont {Tayurskii},\ and\ \citenamefont
  {Volovik}}]{Bernal_graphite_Volovik}%
  \BibitemOpen
  \bibfield  {author} {\bibinfo {author} {\bibfnamefont {P.}~\bibnamefont
  {Esquinazi}}, \bibinfo {author} {\bibfnamefont {T.~T.}\ \bibnamefont
  {Heikkil\"a}}, \bibinfo {author} {\bibfnamefont {Y.~V.}\ \bibnamefont
  {Lysogorskiy}}, \bibinfo {author} {\bibfnamefont {D.~A.}\ \bibnamefont
  {Tayurskii}}, \ and\ \bibinfo {author} {\bibfnamefont {G.~E.}\ \bibnamefont
  {Volovik}},\ }\bibfield  {title} {\enquote {\bibinfo {title} {{On the
  Superconductivity of Graphite Interfaces}},}\ }\href {\doibase
  10.1134/S0021364014170056} {\bibfield  {journal} {\bibinfo  {journal} {JETP
  Lett.}\ }\textbf {\bibinfo {volume} {100}},\ \bibinfo {pages} {336} (\bibinfo
  {year} {2014})}\BibitemShut {NoStop}%
\bibitem [{\citenamefont {Lopes~dos Santos}\ \emph {et~al.}(2007)\citenamefont
  {Lopes~dos Santos}, \citenamefont {Peres},\ and\ \citenamefont
  {Castro~Neto}}]{PhysRevLetT_castro}%
  \BibitemOpen
  \bibfield  {author} {\bibinfo {author} {\bibfnamefont {J.~M.~B.}\
  \bibnamefont {Lopes~dos Santos}}, \bibinfo {author} {\bibfnamefont
  {N.~M.~R.}\ \bibnamefont {Peres}}, \ and\ \bibinfo {author} {\bibfnamefont
  {A.~H.}\ \bibnamefont {Castro~Neto}},\ }\bibfield  {title} {\enquote
  {\bibinfo {title} {{Graphene Bilayer with a Twist: Electronic Structure}},}\
  }\href {\doibase 10.1103/PhysRevLett.99.256802} {\bibfield  {journal}
  {\bibinfo  {journal} {Phys. Rev. Lett.}\ }\textbf {\bibinfo {volume} {99}},\
  \bibinfo {pages} {256802} (\bibinfo {year} {2007})}\BibitemShut {NoStop}%
\bibitem [{\citenamefont {Bistritzer}\ and\ \citenamefont
  {MacDonald}(2011)}]{PNAS_macdonald}%
  \BibitemOpen
  \bibfield  {author} {\bibinfo {author} {\bibfnamefont {R.}~\bibnamefont
  {Bistritzer}}\ and\ \bibinfo {author} {\bibfnamefont {A.~H.}\ \bibnamefont
  {MacDonald}},\ }\bibfield  {title} {\enquote {\bibinfo {title} {{Moir\'e
  bands in twisted double-layer graphene}},}\ }\href {\doibase
  10.1073/pnas.1108174108} {\bibfield  {journal} {\bibinfo  {journal} {PNAS}\
  }\textbf {\bibinfo {volume} {108}},\ \bibinfo {pages} {12233} (\bibinfo
  {year} {2011})}\BibitemShut {NoStop}%
\bibitem [{\citenamefont {D\'ora}\ \emph {et~al.}(2011)\citenamefont {D\'ora},
  \citenamefont {Kailasvuori},\ and\ \citenamefont {Moessner}}]{PhysRevB_Dora}%
  \BibitemOpen
  \bibfield  {author} {\bibinfo {author} {\bibfnamefont {B.}~\bibnamefont
  {D\'ora}}, \bibinfo {author} {\bibfnamefont {J.}~\bibnamefont {Kailasvuori}},
  \ and\ \bibinfo {author} {\bibfnamefont {R.}~\bibnamefont {Moessner}},\
  }\bibfield  {title} {\enquote {\bibinfo {title} {{Lattice generalization of
  the Dirac equation to general spin and the role of the flat band}},}\ }\href
  {\doibase 10.1103/PhysRevB.84.195422} {\bibfield  {journal} {\bibinfo
  {journal} {Phys. Rev. B}\ }\textbf {\bibinfo {volume} {84}},\ \bibinfo
  {pages} {195422} (\bibinfo {year} {2011})}\BibitemShut {NoStop}%
\bibitem [{\citenamefont {Ma\~nes}(2012)}]{PhysRevB_Manes_3fold_theory}%
  \BibitemOpen
  \bibfield  {author} {\bibinfo {author} {\bibfnamefont {J.~L.}\ \bibnamefont
  {Ma\~nes}},\ }\bibfield  {title} {\enquote {\bibinfo {title} {{Existence of
  bulk chiral fermions and crystal symmetry}},}\ }\href {\doibase
  10.1103/PhysRevB.85.155118} {\bibfield  {journal} {\bibinfo  {journal} {Phys.
  Rev. B}\ }\textbf {\bibinfo {volume} {85}},\ \bibinfo {pages} {155118}
  (\bibinfo {year} {2012})}\BibitemShut {NoStop}%
\bibitem [{\citenamefont {Bradlyn}\ \emph {et~al.}(2016)\citenamefont
  {Bradlyn}, \citenamefont {Cano}, \citenamefont {Wang}, \citenamefont
  {Vergniory}, \citenamefont {Felser}, \citenamefont {Cava},\ and\
  \citenamefont {Bernevig}}]{Science_Bernevig_Multifold_theory}%
  \BibitemOpen
  \bibfield  {author} {\bibinfo {author} {\bibfnamefont {B.}~\bibnamefont
  {Bradlyn}}, \bibinfo {author} {\bibfnamefont {J.}~\bibnamefont {Cano}},
  \bibinfo {author} {\bibfnamefont {Z.}~\bibnamefont {Wang}}, \bibinfo {author}
  {\bibfnamefont {M.~G.}\ \bibnamefont {Vergniory}}, \bibinfo {author}
  {\bibfnamefont {C.}~\bibnamefont {Felser}}, \bibinfo {author} {\bibfnamefont
  {R.~J.}\ \bibnamefont {Cava}}, \ and\ \bibinfo {author} {\bibfnamefont
  {B.~A.}\ \bibnamefont {Bernevig}},\ }\bibfield  {title} {\enquote {\bibinfo
  {title} {{Beyond Dirac and Weyl fermions: Unconventional quasiparticles in
  conventional crystals}},}\ }\href {\doibase 10.1126/science.aaf5037}
  {\bibfield  {journal} {\bibinfo  {journal} {Science}\ }\textbf {\bibinfo
  {volume} {353}},\ \bibinfo {pages} {496} (\bibinfo {year}
  {2016})}\BibitemShut {NoStop}%
\bibitem [{\citenamefont {Cao}\ \emph {et~al.}(2018)\citenamefont {Cao},
  \citenamefont {Fatemi}, \citenamefont {Fang}, \citenamefont {Watanabe},
  \citenamefont {Taniguchi}, \citenamefont {Kaxiras},\ and\ \citenamefont
  {Jarillo-Herrero}}]{Exp_TBG_superconductor}%
  \BibitemOpen
  \bibfield  {author} {\bibinfo {author} {\bibfnamefont {Y.}~\bibnamefont
  {Cao}}, \bibinfo {author} {\bibfnamefont {V.}~\bibnamefont {Fatemi}},
  \bibinfo {author} {\bibfnamefont {S.}~\bibnamefont {Fang}}, \bibinfo {author}
  {\bibfnamefont {K.}~\bibnamefont {Watanabe}}, \bibinfo {author}
  {\bibfnamefont {T.}~\bibnamefont {Taniguchi}}, \bibinfo {author}
  {\bibfnamefont {E.}~\bibnamefont {Kaxiras}}, \ and\ \bibinfo {author}
  {\bibfnamefont {P.}~\bibnamefont {Jarillo-Herrero}},\ }\bibfield  {title}
  {\enquote {\bibinfo {title} {{Unconventional superconductivity in magic-angle
  graphene superlattices}},}\ }\href {\doibase 10.1038/nature26160} {\bibfield
  {journal} {\bibinfo  {journal} {Nature}\ }\textbf {\bibinfo {volume} {556}},\
  \bibinfo {pages} {43} (\bibinfo {year} {2018})}\BibitemShut {NoStop}%
\bibitem [{\citenamefont {Zhou}\ \emph {et~al.}()\citenamefont {Zhou},
  \citenamefont {Holleis}, \citenamefont {Saito}, \citenamefont {Cohen},
  \citenamefont {Huynh}, \citenamefont {Patterson}, \citenamefont {Yang},
  \citenamefont {Taniguchi}, \citenamefont {Watanabe},\ and\ \citenamefont
  {Young}}]{AB_stacking_SC}%
  \BibitemOpen
  \bibfield  {author} {\bibinfo {author} {\bibfnamefont {H.}~\bibnamefont
  {Zhou}}, \bibinfo {author} {\bibfnamefont {L.}~\bibnamefont {Holleis}},
  \bibinfo {author} {\bibfnamefont {Y.}~\bibnamefont {Saito}}, \bibinfo
  {author} {\bibfnamefont {L.}~\bibnamefont {Cohen}}, \bibinfo {author}
  {\bibfnamefont {W.}~\bibnamefont {Huynh}}, \bibinfo {author} {\bibfnamefont
  {C.~L.}\ \bibnamefont {Patterson}}, \bibinfo {author} {\bibfnamefont
  {F.}~\bibnamefont {Yang}}, \bibinfo {author} {\bibfnamefont {T.}~\bibnamefont
  {Taniguchi}}, \bibinfo {author} {\bibfnamefont {K.}~\bibnamefont {Watanabe}},
  \ and\ \bibinfo {author} {\bibfnamefont {A.~F.}\ \bibnamefont {Young}},\
  }\bibfield  {title} {\enquote {\bibinfo {title} {{Isospin magnetism and
  spin-triplet superconductivity in Bernal bilayer graphene}},}\ }\href
  {https://arxiv.org/abs/2110.11317} {\bibinfo  {journal} {arxiv:
  arXiv:2110.11317}\ }\BibitemShut {NoStop}%
\bibitem [{\citenamefont {Volovik}(2018)}]{Volovik_flat_band_perspective}%
  \BibitemOpen
\bibfield  {journal} {  }\bibfield  {author} {\bibinfo {author} {\bibfnamefont
  {G.E.}\ \bibnamefont {Volovik}},\ }\bibfield  {title} {\enquote {\bibinfo
  {title} {{Graphite, Graphene, and the Flat Band Superconductivity}},}\ }\href
  {\doibase 10.1134/S0021364018080052} {\bibfield  {journal} {\bibinfo
  {journal} {Jetp Lett.}\ }\textbf {\bibinfo {volume} {107}},\ \bibinfo {pages}
  {516} (\bibinfo {year} {2018})}\BibitemShut {NoStop}%
\bibitem [{\citenamefont {Takane}\ \emph {et~al.}(2019)\citenamefont {Takane},
  \citenamefont {Wang}, \citenamefont {Souma}, \citenamefont {Nakayama},
  \citenamefont {Nakamura}, \citenamefont {Oinuma}, \citenamefont {Nakata},
  \citenamefont {Iwasawa}, \citenamefont {Cacho}, \citenamefont {Kim},
  \citenamefont {Horiba}, \citenamefont {Kumigashira}, \citenamefont
  {Takahashi}, \citenamefont {Ando},\ and\ \citenamefont
  {Sato}}]{PhysRevLett_Takane_Multifold_exp}%
  \BibitemOpen
  \bibfield  {author} {\bibinfo {author} {\bibfnamefont {D.}~\bibnamefont
  {Takane}}, \bibinfo {author} {\bibfnamefont {Z.}~\bibnamefont {Wang}},
  \bibinfo {author} {\bibfnamefont {S.}~\bibnamefont {Souma}}, \bibinfo
  {author} {\bibfnamefont {K.}~\bibnamefont {Nakayama}}, \bibinfo {author}
  {\bibfnamefont {T.}~\bibnamefont {Nakamura}}, \bibinfo {author}
  {\bibfnamefont {H.}~\bibnamefont {Oinuma}}, \bibinfo {author} {\bibfnamefont
  {Y.}~\bibnamefont {Nakata}}, \bibinfo {author} {\bibfnamefont
  {H.}~\bibnamefont {Iwasawa}}, \bibinfo {author} {\bibfnamefont
  {C.}~\bibnamefont {Cacho}}, \bibinfo {author} {\bibfnamefont
  {T.}~\bibnamefont {Kim}}, \bibinfo {author} {\bibfnamefont {K.}~\bibnamefont
  {Horiba}}, \bibinfo {author} {\bibfnamefont {H.}~\bibnamefont {Kumigashira}},
  \bibinfo {author} {\bibfnamefont {T.}~\bibnamefont {Takahashi}}, \bibinfo
  {author} {\bibfnamefont {Y.}~\bibnamefont {Ando}}, \ and\ \bibinfo {author}
  {\bibfnamefont {T.}~\bibnamefont {Sato}},\ }\bibfield  {title} {\enquote
  {\bibinfo {title} {{Observation of Chiral Fermions with a Large Topological
  Charge and Associated Fermi-Arc Surface States in CoSi}},}\ }\href {\doibase
  10.1103/PhysRevLett.122.076402} {\bibfield  {journal} {\bibinfo  {journal}
  {Phys. Rev. Lett.}\ }\textbf {\bibinfo {volume} {122}},\ \bibinfo {pages}
  {076402} (\bibinfo {year} {2019})}\BibitemShut {NoStop}%
\bibitem [{\citenamefont {Rao}\ \emph {et~al.}(2019)\citenamefont {Rao},
  \citenamefont {Li}, \citenamefont {Zhang}, \citenamefont {Tian},
  \citenamefont {Li}, \citenamefont {Fu}, \citenamefont {Tang}, \citenamefont
  {Wang}, \citenamefont {Li}, \citenamefont {Fan}, \citenamefont {Li},
  \citenamefont {Huang}, \citenamefont {Liu}, \citenamefont {Long},
  \citenamefont {Fang}, \citenamefont {Weng}, \citenamefont {Shi},
  \citenamefont {Lei}, \citenamefont {Sun}, \citenamefont {Qian},\ and\
  \citenamefont {Ding}}]{Nature_Zhicheng_Multifold_exp}%
  \BibitemOpen
  \bibfield  {author} {\bibinfo {author} {\bibfnamefont {Z.}~\bibnamefont
  {Rao}}, \bibinfo {author} {\bibfnamefont {H.}~\bibnamefont {Li}}, \bibinfo
  {author} {\bibfnamefont {T.}~\bibnamefont {Zhang}}, \bibinfo {author}
  {\bibfnamefont {S.}~\bibnamefont {Tian}}, \bibinfo {author} {\bibfnamefont
  {C.}~\bibnamefont {Li}}, \bibinfo {author} {\bibfnamefont {B.}~\bibnamefont
  {Fu}}, \bibinfo {author} {\bibfnamefont {C.}~\bibnamefont {Tang}}, \bibinfo
  {author} {\bibfnamefont {L.}~\bibnamefont {Wang}}, \bibinfo {author}
  {\bibfnamefont {Z.}~\bibnamefont {Li}}, \bibinfo {author} {\bibfnamefont
  {W.}~\bibnamefont {Fan}}, \bibinfo {author} {\bibfnamefont {J.}~\bibnamefont
  {Li}}, \bibinfo {author} {\bibfnamefont {Y.}~\bibnamefont {Huang}}, \bibinfo
  {author} {\bibfnamefont {Z.}~\bibnamefont {Liu}}, \bibinfo {author}
  {\bibfnamefont {Y.}~\bibnamefont {Long}}, \bibinfo {author} {\bibfnamefont
  {C.}~\bibnamefont {Fang}}, \bibinfo {author} {\bibfnamefont {H.}~\bibnamefont
  {Weng}}, \bibinfo {author} {\bibfnamefont {Y.}~\bibnamefont {Shi}}, \bibinfo
  {author} {\bibfnamefont {H.}~\bibnamefont {Lei}}, \bibinfo {author}
  {\bibfnamefont {Y.}~\bibnamefont {Sun}}, \bibinfo {author} {\bibfnamefont
  {T.}~\bibnamefont {Qian}}, \ and\ \bibinfo {author} {\bibfnamefont
  {H.}~\bibnamefont {Ding}},\ }\bibfield  {title} {\enquote {\bibinfo {title}
  {{Topological chiral crystals with helicoid-arc quantum states}},}\ }\href
  {\doibase 10.1038/s41586-019-1031-8} {\bibfield  {journal} {\bibinfo
  {journal} {Nature}\ }\textbf {\bibinfo {volume} {567}},\ \bibinfo {pages}
  {496} (\bibinfo {year} {2019})}\BibitemShut {NoStop}%
\bibitem [{\citenamefont {Sanchez}\ \emph {et~al.}(2019)\citenamefont
  {Sanchez}, \citenamefont {Belopolski}, \citenamefont {Cochran}, \citenamefont
  {Xu}, \citenamefont {Yin}, \citenamefont {Chang}, \citenamefont {Xie},
  \citenamefont {Manna}, \citenamefont {S\"uß}, \citenamefont {Huang},
  \citenamefont {Alidoust}, \citenamefont {Multer}, \citenamefont {Zhang},
  \citenamefont {Shumiya}, \citenamefont {Wang}, \citenamefont {Wang},
  \citenamefont {Chang}, \citenamefont {Felser}, \citenamefont {Xu},
  \citenamefont {Jia}, \citenamefont {Lin},\ and\ \citenamefont
  {Hasan}}]{Nature_Sanchez_Multifold_exp}%
  \BibitemOpen
  \bibfield  {author} {\bibinfo {author} {\bibfnamefont {D.~S.}\ \bibnamefont
  {Sanchez}}, \bibinfo {author} {\bibfnamefont {I.}~\bibnamefont {Belopolski}},
  \bibinfo {author} {\bibfnamefont {T.~A.}\ \bibnamefont {Cochran}}, \bibinfo
  {author} {\bibfnamefont {X.}~\bibnamefont {Xu}}, \bibinfo {author}
  {\bibfnamefont {J.-X.}\ \bibnamefont {Yin}}, \bibinfo {author} {\bibfnamefont
  {G.}~\bibnamefont {Chang}}, \bibinfo {author} {\bibfnamefont
  {W.}~\bibnamefont {Xie}}, \bibinfo {author} {\bibfnamefont {K.}~\bibnamefont
  {Manna}}, \bibinfo {author} {\bibfnamefont {V.}~\bibnamefont {S\"uß}},
  \bibinfo {author} {\bibfnamefont {C.-Y.}\ \bibnamefont {Huang}}, \bibinfo
  {author} {\bibfnamefont {N.}~\bibnamefont {Alidoust}}, \bibinfo {author}
  {\bibfnamefont {D.}~\bibnamefont {Multer}}, \bibinfo {author} {\bibfnamefont
  {S.~S.}\ \bibnamefont {Zhang}}, \bibinfo {author} {\bibfnamefont
  {N.}~\bibnamefont {Shumiya}}, \bibinfo {author} {\bibfnamefont
  {X.}~\bibnamefont {Wang}}, \bibinfo {author} {\bibfnamefont {G.-Q.}\
  \bibnamefont {Wang}}, \bibinfo {author} {\bibfnamefont {T.-R.}\ \bibnamefont
  {Chang}}, \bibinfo {author} {\bibfnamefont {C.}~\bibnamefont {Felser}},
  \bibinfo {author} {\bibfnamefont {S.-Y.}\ \bibnamefont {Xu}}, \bibinfo
  {author} {\bibfnamefont {S.}~\bibnamefont {Jia}}, \bibinfo {author}
  {\bibfnamefont {H.}~\bibnamefont {Lin}}, \ and\ \bibinfo {author}
  {\bibfnamefont {M.~Z.}\ \bibnamefont {Hasan}},\ }\bibfield  {title} {\enquote
  {\bibinfo {title} {{Topological chiral crystals with helicoid-arc quantum
  states}},}\ }\href {\doibase 10.1038/s41586-019-1037-2} {\bibfield  {journal}
  {\bibinfo  {journal} {Nature}\ }\textbf {\bibinfo {volume} {567}},\ \bibinfo
  {pages} {500} (\bibinfo {year} {2019})}\BibitemShut {NoStop}%
\bibitem [{\citenamefont {Lv}\ \emph {et~al.}(2021)\citenamefont {Lv},
  \citenamefont {Qian},\ and\ \citenamefont {Ding}}]{RevModPhys_Ding}%
  \BibitemOpen
  \bibfield  {author} {\bibinfo {author} {\bibfnamefont {B.~Q.}\ \bibnamefont
  {Lv}}, \bibinfo {author} {\bibfnamefont {T.}~\bibnamefont {Qian}}, \ and\
  \bibinfo {author} {\bibfnamefont {H.}~\bibnamefont {Ding}},\ }\bibfield
  {title} {\enquote {\bibinfo {title} {{Experimental perspective on
  three-dimensional topological semimetals}},}\ }\href {\doibase
  10.1103/RevModPhys.93.025002} {\bibfield  {journal} {\bibinfo  {journal}
  {Rev. Mod. Phys.}\ }\textbf {\bibinfo {volume} {93}},\ \bibinfo {pages}
  {025002} (\bibinfo {year} {2021})}\BibitemShut {NoStop}%
\bibitem [{\citenamefont {Lin}(2020)}]{PhysRevResearch_Lin}%
  \BibitemOpen
  \bibfield  {author} {\bibinfo {author} {\bibfnamefont {Yu-Ping}\ \bibnamefont
  {Lin}},\ }\bibfield  {title} {\enquote {\bibinfo {title} {{Chiral flat band
  superconductivity from symmetry-protected three-band crossings}},}\ }\href
  {\doibase 10.1103/PhysRevResearch.2.043209} {\bibfield  {journal} {\bibinfo
  {journal} {Phys. Rev. Research}\ }\textbf {\bibinfo {volume} {2}},\ \bibinfo
  {pages} {043209} (\bibinfo {year} {2020})}\BibitemShut {NoStop}%
\bibitem [{\citenamefont {Sac\'ep\'e}\ \emph {et~al.}(2020)\citenamefont
  {Sac\'ep\'e}, \citenamefont {Feigel'man},\ and\ \citenamefont
  {Klapwijk}}]{Nature_Sacepe_Feigelman}%
  \BibitemOpen
  \bibfield  {author} {\bibinfo {author} {\bibfnamefont {B.}~\bibnamefont
  {Sac\'ep\'e}}, \bibinfo {author} {\bibfnamefont {M.}~\bibnamefont
  {Feigel'man}}, \ and\ \bibinfo {author} {\bibfnamefont {T.~M.}\ \bibnamefont
  {Klapwijk}},\ }\bibfield  {title} {\enquote {\bibinfo {title} {{Quantum
  breakdown of superconductivity in low-dimensional materials}},}\ }\href
  {\doibase 10.1038/s41567-020-0905-x} {\bibfield  {journal} {\bibinfo
  {journal} {Nat. Phys.}\ }\textbf {\bibinfo {volume} {16}},\ \bibinfo {pages}
  {734} (\bibinfo {year} {2020})}\BibitemShut {NoStop}%
\bibitem [{\citenamefont {Peotta}\ and\ \citenamefont
  {T\"orm\"a}(2015)}]{Peotta_Torma}%
  \BibitemOpen
  \bibfield  {author} {\bibinfo {author} {\bibfnamefont {S.}~\bibnamefont
  {Peotta}}\ and\ \bibinfo {author} {\bibfnamefont {P.}~\bibnamefont
  {T\"orm\"a}},\ }\bibfield  {title} {\enquote {\bibinfo {title}
  {{Superfluidity in topologically nontrivial flat bands}},}\ }\href {\doibase
  10.1038/ncomms9944} {\bibfield  {journal} {\bibinfo  {journal} {Nat.
  Commun.}\ }\textbf {\bibinfo {volume} {6}},\ \bibinfo {pages} {8944}
  (\bibinfo {year} {2015})}\BibitemShut {NoStop}%
\bibitem [{\citenamefont {Nozi\`eres}\ and\ \citenamefont
  {Schmitt-Rink}(1985)}]{JLTP_Schmitt}%
  \BibitemOpen
  \bibfield  {author} {\bibinfo {author} {\bibfnamefont {P.}~\bibnamefont
  {Nozi\`eres}}\ and\ \bibinfo {author} {\bibfnamefont {S.}~\bibnamefont
  {Schmitt-Rink}},\ }\bibfield  {title} {\enquote {\bibinfo {title} {{Bose
  condensation in an attractive fermion gas: From weak to strong coupling
  superconductivity}},}\ }\href {\doibase 10.1007/BF00683774} {\bibfield
  {journal} {\bibinfo  {journal} {J. Low Temp. Phys.}\ }\textbf {\bibinfo
  {volume} {59}},\ \bibinfo {pages} {195} (\bibinfo {year} {1985})}\BibitemShut
  {NoStop}%
\bibitem [{\citenamefont {Zyuzin}()}]{Zyuzin_preformed}%
  \BibitemOpen
  \bibfield  {author} {\bibinfo {author} {\bibfnamefont {A.~Yu.}\ \bibnamefont
  {Zyuzin}},\ }\bibfield  {title} {\enquote {\bibinfo {title}
  {{Superconductivity in dilute system of sites with strong electron-electron
  attraction}},}\ }\href {https://arxiv.org/abs/2012.12597} {\bibinfo
  {journal} {arxiv: arXiv:2012.12597}\ }\BibitemShut {NoStop}%
\bibitem [{SM_()}]{SM_editors}%
  \BibitemOpen
\bibfield  {journal} {  }\href@noop {} {}\bibinfo {note} {{See Supplemental
  Material for Green's function, interaction in the Cooper channel, and
  mean-field approach to preformed-pair to phase-coherent
  transition.}}\BibitemShut {Stop}%
\bibitem [{\citenamefont {Mandal}\ \emph {et~al.}(2021)\citenamefont {Mandal},
  \citenamefont {Link},\ and\ \citenamefont {Herbut}}]{Herbut}%
  \BibitemOpen
  \bibfield  {author} {\bibinfo {author} {\bibfnamefont {S.}~\bibnamefont
  {Mandal}}, \bibinfo {author} {\bibfnamefont {J.~M.}\ \bibnamefont {Link}}, \
  and\ \bibinfo {author} {\bibfnamefont {I.~F.}\ \bibnamefont {Herbut}},\
  }\bibfield  {title} {\enquote {\bibinfo {title} {Time-reversal symmetry
  breaking and $d$-wave superconductivity of triple-point fermions},}\ }\href
  {\doibase 10.1103/PhysRevB.104.134512} {\bibfield  {journal} {\bibinfo
  {journal} {Phys. Rev. B}\ }\textbf {\bibinfo {volume} {104}},\ \bibinfo
  {pages} {134512} (\bibinfo {year} {2021})}\BibitemShut {NoStop}%
\bibitem [{\citenamefont {Abrikosov}\ \emph {et~al.}(1975)\citenamefont
  {Abrikosov}, \citenamefont {Gorkov},\ and\ \citenamefont
  {Dzyaloshinski}}]{AGD}%
  \BibitemOpen
  \bibfield  {author} {\bibinfo {author} {\bibfnamefont {A.~A.}\ \bibnamefont
  {Abrikosov}}, \bibinfo {author} {\bibfnamefont {L.~P.}\ \bibnamefont
  {Gorkov}}, \ and\ \bibinfo {author} {\bibfnamefont {I.~E.}\ \bibnamefont
  {Dzyaloshinski}},\ }\href@noop {} {\emph {\bibinfo {title} {{Methods of
  quantum field theory in statistical physics}}}}\ (\bibinfo  {publisher}
  {Dover publications},\ \bibinfo {year} {1975})\ \bibinfo {note} {ch. 7, sec.
  38}\BibitemShut {NoStop}%
\bibitem [{\citenamefont {Song}\ and\ \citenamefont
  {Bernevig}()}]{Bernevig_flat_band}%
  \BibitemOpen
  \bibfield  {author} {\bibinfo {author} {\bibfnamefont {Z.-D.}\ \bibnamefont
  {Song}}\ and\ \bibinfo {author} {\bibfnamefont {B.~A.}\ \bibnamefont
  {Bernevig}},\ }\bibfield  {title} {\enquote {\bibinfo {title} {{MATBG as
  Topological Heavy Fermion: I. Exact Mapping and Correlated Insulators}},}\
  }\href {https://arxiv.org/abs/2111.05865} {\bibinfo  {journal} {arxiv:
  arXiv:2111.05865}\ }\BibitemShut {NoStop}%
\end{thebibliography}%

\clearpage
\onecolumngrid
\begin{center}
\rule{0.38\linewidth}{1pt}\\
\vspace{-0.37cm}\rule{0.49\linewidth}{1pt}
\end{center}
\setcounter{section}{0}
\setcounter{equation}{0}

\section*{Supplemental Material to \texorpdfstring{\\}{}
"Preformed Cooper pairs in flat-band semimetals"}

\section{Model}
Here we present more details of the superconductivity in flat-band materials. Consider a band structure consisting of two valleys, in which
a flat band intersects with a Dirac point. So that each valley hosts three bands. Assume that the Dirac points are positioned at momenta $\pm \mathbf{K}_D$. 
Consider a low-energy model of a semimetal with a pair of such points. We ignore the single-particle inter-valley tunneling processes. The model Hamiltonian can be represented via a sum of two independent contributions from two valleys 
[\ref{Bradlyn_SM}]. 
\begin{equation}\label{first_eq}
\mathcal{H} = \int_{\mathbf{q}} \Psi^{\dag}_{+,\mathbf{q}} v_{\mathrm{F}}\mathbf{S}\cdot (\mathbf{q}- \mathbf{K}_\mathrm{D}) \Psi_{+,\mathbf{q}} + 
\int_{\mathbf{q}} \Psi^{\dag}_{-,\mathbf{q}} v_{\mathrm{F}}\mathbf{S}\cdot(\mathbf{q} + \mathbf{K}_\mathrm{D}) \Psi_{-,\mathbf{q}}, 
\end{equation}
where $v_{\mathrm{F}}$ is the Fermi velocity, $\int_{\bf q}(..) \equiv \int \frac{d\mathbf{q}}{(2\pi)^3}(..)$. It is convenient to rewrite (\ref{first_eq}) as
\begin{equation}\label{SM_nonint}
\mathcal{H} = \int_{\mathbf{k}} \sum_{s=\pm}\Psi^{\dag}_{s,\mathbf{k}} v_{\mathrm{F}}\mathbf{S}\cdot \mathbf{k} \Psi_{s,\mathbf{k}}, 
\end{equation}
where now $\mathbf{k}$ is the momentum measured relatively to the $\pm \mathbf{K}_{\mathrm{D}}$ with $k \ll K_{\mathrm{D}}$.
The electron operators are defined by 
\begin{equation}
\Psi_{s,\mathbf{k}} = [\Psi_{s, +1,\mathbf{k}}, \Psi_{s, 0,\mathbf{k}}, \Psi_{s, -1,\mathbf{k}}]^{T}, 
\end{equation}
where indices $\pm 1, 0$ correspond to three different bands,
two of which are dispersive, $E_{\pm 1} = \pm v_{\mathrm{F}} k$, and another is flat, $E_{ 0} = 0$. The latter is considered in 
the infinite mass limit approximation, so that higher order momentum corrections are neglected. We will be using $\hbar = k_{\mathrm{B}} = 1$ units throughout the paper.
Finally, $\mathbf{S} = (S_x,S_y,S_z)$ are the Gell-Mann matrices acting on "which band" pseudospin degree of freedom:
\begin{align}
\begin{gathered}
S_x= \frac{1}{\sqrt{2}}\left(
 \begin{matrix}
 0& 1& 0 \\
 1& 0& 1\\
 0& 1& 0
 \end{matrix}
\right), ~
S_y= \frac{i}{\sqrt{2}}\left(
 \begin{matrix}
 0& -1& 0 \\
 1& 0& -1\\
 0& 1& 0
 \end{matrix}
\right),
S_z= \left(
 \begin{matrix}
 1& 0& 0 \\
 0& 0& 0\\
 0& 0& -1
 \end{matrix}
\right),
S_0= \left(
 \begin{matrix}
 1& 0& 0 \\
 0& 1& 0\\
 0& 0& 1
 \end{matrix}
\right)\equiv 1.
\end{gathered}
\end{align}
The unit matrix will be used when needed to avoid confusion. Useful identities: $S_x^2+S_y^2+S_z^2=2S_0$, $S_i = S_i^{\dag}$, $[S_a,S_b]=i \varepsilon_{abc}S_c$.

%%%%%%%%%%%%%%%%%%%%%% BEGIN FIGURE %%%%%%%%%%%%%%%%%%%
\begin{figure}[h]
\centering
\includegraphics[width=5.0cm]{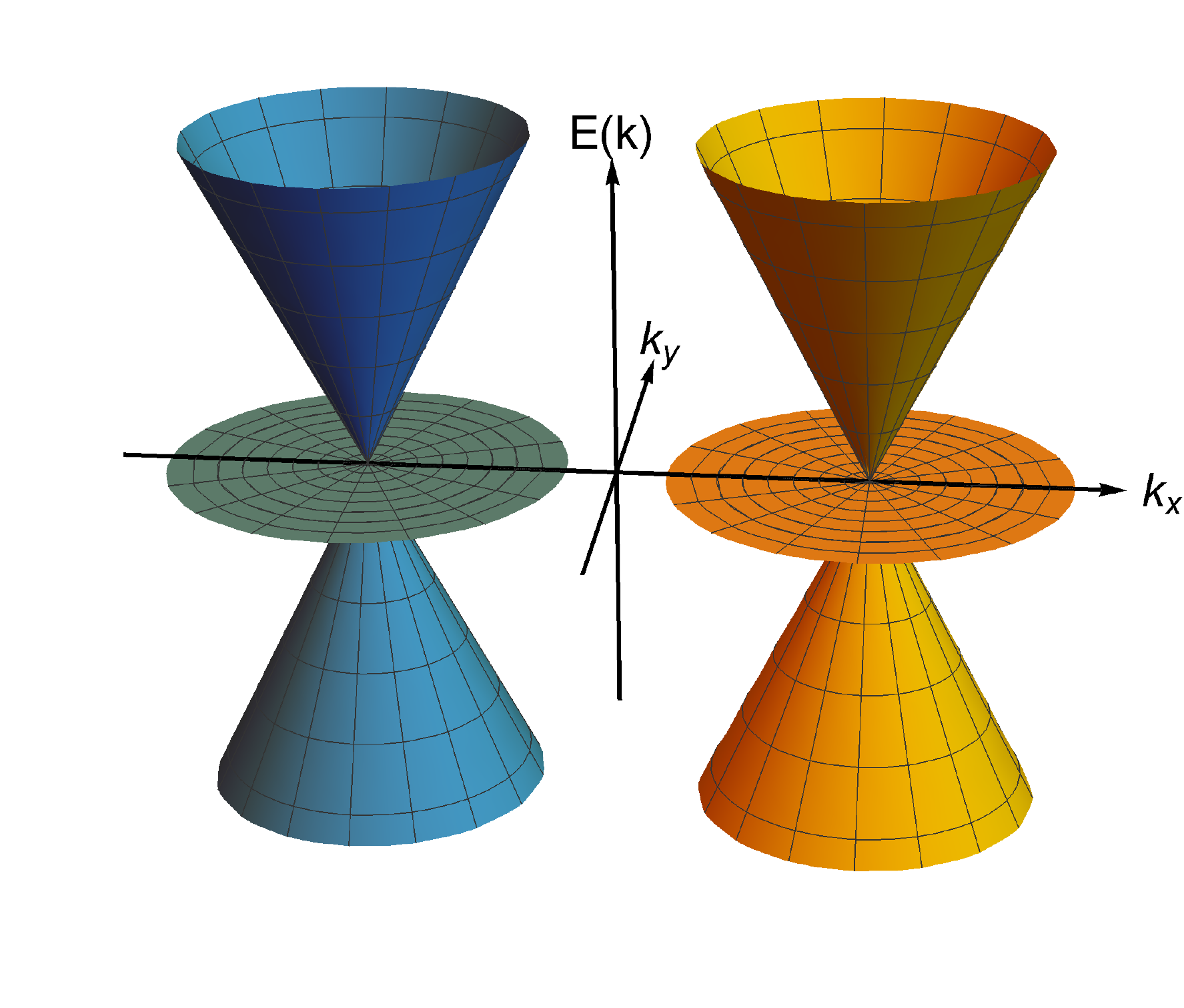}
\caption{\label{fig1_SM} Schematics of the band structure $E(\mathbf{k})$ in the vicinity of two three-band-touching points. 
There are two points at which Dirac cones and flat bands intersect.}
\end{figure}
%%%%%%%%%%%%%%%%%%%%%% END FIGURE %%%%%%%%%%%%%%%%%%%

\subsection{Green function in momentum space}
Let us first write down the Green function of electrons [\ref{AGD_book_SM}] for valleys $\pm\mathbf{K}_\mathrm{D}$.
\begin{equation}\label{GF_1_SM}
G(\mathbf{k},i\omega_n) = [(i\omega_n+\mu)S_0 - v_{\mathrm{F}}\mathbf{k}\cdot \mathbf{S}]^{-1} = \frac{S_0- (\mathbf{S}\mathbf{n}_{k})^2}{i\omega_n +\mu}
+ \frac{1}{2}\sum_{s=\pm1} \frac{ (\mathbf{S}\mathbf{n}_k)^2 + s(\mathbf{S}\mathbf{n}_k)}{i\omega_n + \mu - sv_{\mathrm{F}} k}.
\end{equation}
Here $\mathbf{n}_k = {\bf k}/k$ is the unit vector in the direction of $\mathbf{k}$, and $\omega_n = (2n+1)\pi T$ is the fermionic Matsubara frequency and $T$ is the temperature. 
We will keep the chemical potential $\mu \geq 0$ in what follows.
It is instructive to revisit the electron density of states in this system. At frequency $\omega$ per one valley it is defined by
\begin{equation}
\nu_{\mathrm{tot}}(\omega)  = -\frac{i}{\pi}\mathrm{Tr} \int_{\mathbf{k}} \mathrm{Im} G(\mathbf{k},\omega +i\delta). 
\end{equation}
Using
\begin{equation}
\int_{\mathbf{k}}[S_0- (\mathbf{S}\mathbf{n}_{k})^2] = S_0\frac{K^3}{3},~~~ \boxed{K^3 \equiv \int \frac{d^3k}{(2\pi)^3}}
\end{equation}
where $K^3$ is defined by the volume of the flat band in momentum space, one obtains
\begin{equation}
\nu_{\mathrm{tot}}(\omega) = K^3 \delta(\omega +\mu)
+ \frac{(\omega+\mu)^2}{2\pi^2 v_{\mathrm{F}}^3}\equiv \nu_{\mathrm{loc}}(\omega) +\nu_{\mathrm{nl}}(\omega).
\end{equation}
The delta-function term originates from the flat-band contribution. The second term is coming from the Dirac-band contribution.

\subsection{Locality and non-locality seen via Green function}
Let us write down Green function in position-frequency representation
\begin{eqnarray}\nonumber
G(\mathbf{r},i\omega_n) &=& \int_{\mathbf{k}}G(\mathbf{k},i\omega_n) e^{i\mathbf{k r}}  = \frac{1}{i\omega_n +\mu}\left\{S_0\delta(\mathbf{r})+  (\mathbf{S}\boldsymbol{\partial}_r)^2 \frac{1}{4\pi r} \right\} 
\\
&+& i \mathbf{S}\boldsymbol{\partial}_r \frac{1}{4\pi v_{\mathrm{F}} r} e^{-\frac{r}{v}(|\omega_n|-i\mu \mathrm{sgn}\omega_n)}
-
\frac{(\mathbf{S}\boldsymbol{\partial}_r)^2}{i\omega_n +\mu} \frac{1}{4\pi r}
\left[1-e^{-\frac{r}{v_{\mathrm{F}}}(|\omega_n|-i\mu \mathrm{sgn}\omega_n)}\right] 
\equiv G_{\mathrm{nl}}(\mathbf{r},i\omega_n)  +G_{\mathrm{loc}}(\mathbf{r},i\omega_n) .
\end{eqnarray}
We identify that flat-band and Dirac band contributions as local and nonlocal, respectively. As we will see below, the Dirac band gives small correction to the local flat-band contribution. 
Indeed, the flat-band contribution to the Green function reads
\begin{eqnarray}\label{GF_local_SM}
G_{\mathrm{loc}}(\mathbf{r},i\omega_n) = \frac{1}{i\omega_n +\mu}\left\{S_0 \delta(\mathbf{r})+ \frac{1}{4\pi r^3}[3(\mathbf{S}\mathbf{n}_r)^2 - \mathbf{S}^2]\right\}.
\end{eqnarray}
Here $\mathbf{n}_r = \mathbf{r}/r$ is the unit vector in the direction of $\mathbf{r}$. We see that the probability amplitude for the fermion to propagate is localized. 
The frequency and spatial dependent terms are decoupled in contrast to the Dirac band Green function part.

The Dirac band contribution to the Green function in spatial coordinate representation is given by
\begin{eqnarray}
G_{\mathrm{nl}}(\mathbf{r},i\omega_n) = -\frac{i(\mathbf{S}\mathbf{n}_r)}{4\pi v_{\mathrm{F}} r^2}\left(1+\frac{r}{L_{\omega}}\right)e^{-\frac{r}{L_{\omega}}} -
\frac{iL_{\omega} \mathrm{sgn}\omega_n}{4\pi v_{\mathrm{F}} r^3} \left\{[\mathbf{S}^2-3(\mathbf{S}\mathbf{n}_r)^2] \left[1-\left(1-\frac{r}{L_{\omega}}\right)e^{-\frac{r}{L_{\omega}}}\right] 
+ (\mathbf{S}\mathbf{n}_r)^2\frac{r^2}{L^2_{\omega}}e^{-\frac{r}{L_{\omega}}}\right\},~~~~~
\end{eqnarray}
where $L^{-1}_{\omega} = v_{\mathrm{F}}^{-1}(\omega_n-i\mu )\mathrm{sgn}\omega_n$ is introduced for brevity. We note that in the limit of $r\rightarrow 0$, the Green function is cut by the inter-atomic distance. 
At small lengths, we estimate
\begin{equation}
G_{\mathrm{nl}}(\mathbf{r},i\omega_n)|_{r \rightarrow 0} \rightarrow - \frac{i\mathrm{sgn}\omega_n}{\pi v_{\mathrm{F}}r^2}.  
\end{equation}
The spatial dependence at small lengths is weaker compared with the dipole-like term in expression (\ref{GF_local_SM}).

On the other hand, at large chemical potential $\mu \gg |\omega_n|$ for lengths larger than the Fermi wave-length, $\mu r/v_{\mathrm{F}} \gg 1$, we obtain
\begin{eqnarray}
G_{\mathrm{nl}}(\mathbf{r}, i\omega_n) = -\frac{\mu }{4\pi v_{\mathrm{F}}^2 r}  [(\mathbf{S}\mathbf{n}_r) \mathrm{sgn}\omega_n + (\mathbf{S}\mathbf{n}_r)^2 ]e^{-\frac{r}{v_{\mathrm{F}}}(\omega_n-i\mu) \mathrm{sgn}\omega_n}.~~~
\end{eqnarray}
The Green function has standard spatial dependence in 3D and unusual pseudospin matrices dependence compared with the spin-1/2 case.
Here the Green function is cut by the Fermi wave-length in the limit of $r\rightarrow 0$. 

\section{Model of Cooper pairing}

Consider s-wave Cooper pairing, in which a Cooper pair is formed by two electrons from different valleys. Note that the pseudospin of an electron is one. Hence, the total pseudospin of a Cooper pair can take values
$S = 0,1,2$. The order parameter in the channel with total pseudospin $S$ can have $2S+1$ components. It comes from counting the $z$-projections of the Cooper pair's total pseudospin.
There is one component in the $S=0$ case. There are three and five components in the $S=1$ and $S=2$ channels, respectively. 
Due to Pauli principle, the $s$-wave inter-valley-odd pairing can have total pseudospin of a Cooper pair $S = 0$ and $S=2$. The Cooper pair wave-function has pseudospin-even symmetry.
Note that the property of the channel is different compared with the spin-1/2 quasiparticles.
In conventional $s$-wave superconductors the Cooper pair wave-function has pseudospin-odd symmetry.

%%%%%%%%%%%%%%%%%%%%%% BEGIN FIGURE %%%%%%%%%%%%%%%%%%%
\begin{figure}[h]
\centering
\includegraphics[width=7.0cm]{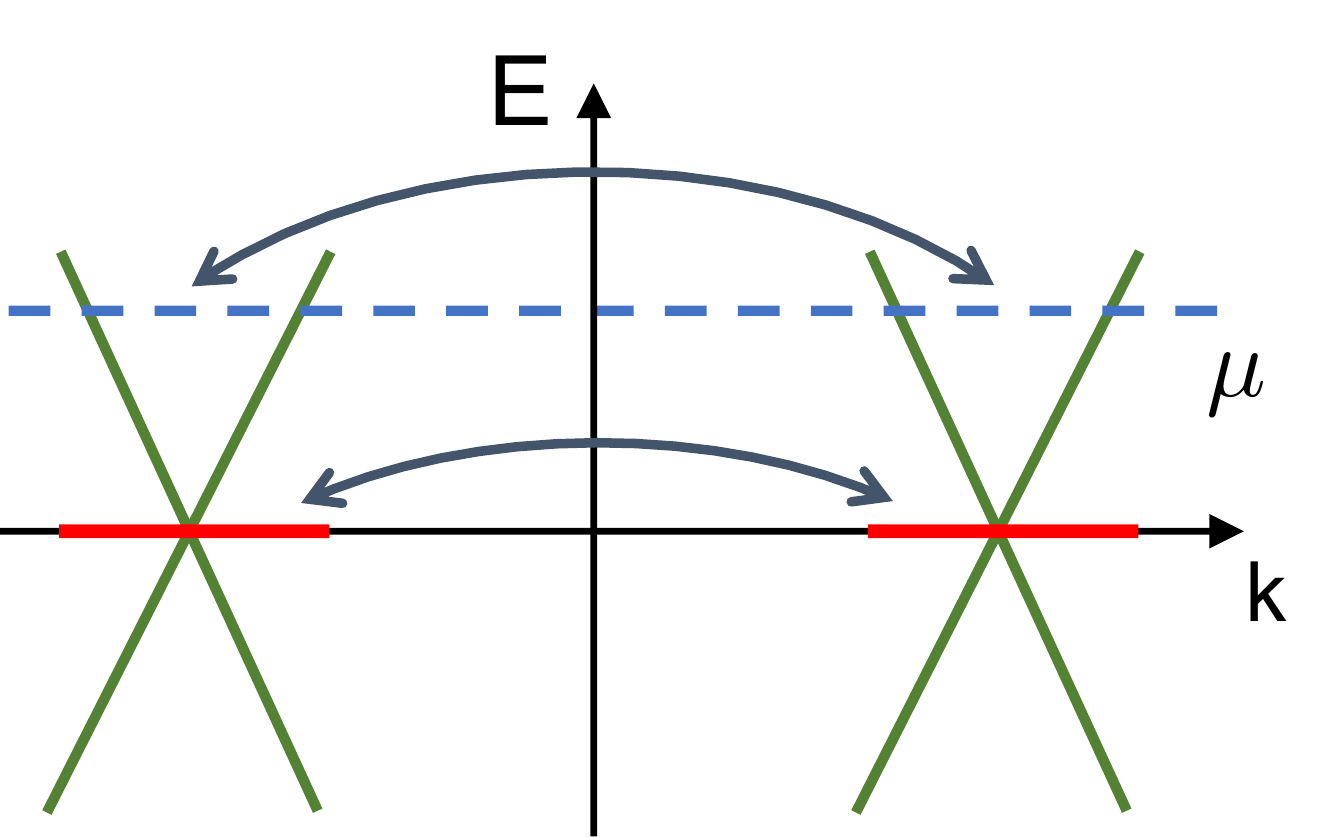}
\caption{\label{fig1_SM} Schematics of the band structure $E(\mathbf{k})$ in the vicinity of two three-band-touching points at chemical potential $\mu$. 
There are two points at which Dirac cones and flat bands intersect. Superconducting pairing of electrons from different valleys is considered.}
\end{figure}
%%%%%%%%%%%%%%%%%%%%%% END FIGURE %%%%%%%%%%%%%%%%%%%

\subsection{Symmetry}
To proceed with the interaction in the Cooper channel, we shall introduce matrices for different irreducible pseudospin-1 representations.
To construct these matrices, we consider the eigenfunctions of operator $S_z$:
\begin{equation}
\psi_{+}
=
\left(
 \begin{matrix}
  1\\
  0\\
   0
 \end{matrix}
\right),
\psi_{0}
=
\left(
 \begin{matrix}
  0\\
  1\\
   0
 \end{matrix}
\right),
\psi_{-}
=
\left(
 \begin{matrix}
  0\\
  0\\
   1
 \end{matrix}
\right).
\end{equation}
For pseudospin-1 particles, it is convenient to introduce $\mathbf{M}_{S}\gamma = (M_{S,S}\gamma, M_{S,S-1}\gamma,..., M_{S,-S}\gamma)$, following Ref.~[\ref{Nandkishore}]. 
The components are given by
\begin{align}
&\mathbf{M}_{0} \equiv M_{0,0}= S_0,\\
&\mathbf{M}_1= (M_{1,1},M_{1,0},M_{1,-1})= \frac{\sqrt{3}}{2} \left(- S_+, \sqrt{2} S_z, S_- \right),\\
&\mathbf{M}_2 = (M_{2,2},M_{2,1},M_{2,0},M_{2,-1},M_{2,-2}) =\frac{\sqrt{3}}{2}\left(S_+^2, -\{S_+,S_z\}, \sqrt{\frac{2}{3}} \left[3S_z^2-\mathbf{S}^2\right], \{S_-,S_z\}, S_-^2 \right).
\end{align}
where $S_{\pm} = S_x+i S_y$. Useful identities 
\begin{align}
&\mathbf{M}_0\cdot \mathbf{M}^{\dag}_0 = S_0,~~ \mathbf{M}_1\cdot \mathbf{M}^{\dag}_1 = 3S_0,~~ \mathbf{M}_2\cdot \mathbf{M}^{\dag}_2 = 5S_0, \\
&\textrm{Tr}(M_{S,m_S} M^{\dag}_{S',m'_{S'}}) = 3\delta_{SS'}\delta_{m_Sm'_{S'}},
\end{align}
where $m_S= -S,-S+1,...,S-1,S$. And it is convenient to introduce a unitary operator 
\begin{equation}
\gamma = e^{i\pi S_y}=
\left(
 \begin{matrix}
  0&0&1\\
  0&-1&0\\
   1&0&0
 \end{matrix}
\right),
\end{equation}
which satisfies $\gamma^2=1, ~\gamma\gamma^{\dag}=1$ and transforms the pseudospin-1 operators as 
\begin{equation}
\gamma \mathbf{S}^* \gamma = - \mathbf{S},~~~ \gamma \mathbf{S}_{\pm} \gamma = - \mathbf{S}_{\mp},
\end{equation} 
and 
\begin{equation}
S_x\gamma = \frac{1}{\sqrt{2}}
\left(
 \begin{matrix}
  0&-1&0\\
  1&0&1\\
   0&-1&0
 \end{matrix}
\right),~~
S_y\gamma = \frac{i}{\sqrt{2}}
\left(
 \begin{matrix}
  0&1&0\\
  -1&0&1\\
   0&-1&0
 \end{matrix}
\right),~~
S_z\gamma =
\left(
 \begin{matrix}
  0&0&1\\
  0&0&0\\
   -1&0&0
 \end{matrix}
\right) .
\end{equation}
Note that $\gamma$ resembles the antisymmetric spin-matrix structure of the gap function in usual superconductors.

\subsubsection{Total spin $S=0$ of the Cooper pair} 
This case is trivial.
\begin{equation}
M_{0,0} = \left(
 \begin{matrix}
  1&0&0\\
  0&1&0\\
   0&0&1
 \end{matrix}
\right),~~~ M_{0,0}\gamma = \left(
 \begin{matrix}
  0&0&1\\
  0&-1&0\\
   1&0&0
 \end{matrix}
\right). 
\end{equation}

\subsubsection{Total spin $S=1$ of the Cooper pair} 
Here are three components.
\begin{equation}
M_{1,1}
=
\sqrt{\frac{3}{2}} \left(
 \begin{matrix}
  0&-1&0\\
  0&0&-1\\
   0&0&0
 \end{matrix}
\right),~~
M_{1,0}
=
\sqrt{\frac{3}{2}} \left(
 \begin{matrix}
  1&0&0\\
  0&0&0\\
   0&0&-1
 \end{matrix}
\right)~~
M_{1,-1}=
\sqrt{\frac{3}{2}} \left(
 \begin{matrix}
  0&0&0\\
  1&0&0\\
   0&1&0
 \end{matrix}
\right).
\end{equation}

It is convenient to write down explicitly:
\begin{equation}
M_{1,1}\gamma =
\sqrt{\frac{3}{2}}
\left(
 \begin{matrix}
  0&1&0\\
  -1&0&0\\
   0&0&0
 \end{matrix}
\right),~~
M_{1,0}\gamma
=
\sqrt{\frac{3}{2}} \left(
 \begin{matrix}
  0&0&1\\
  0&0&0\\
   -1&0&0
 \end{matrix}
\right)~~
M_{1,-1}\gamma =
\sqrt{\frac{3}{2}}\left(
 \begin{matrix}
  0&0&0\\
  0&0&1\\
   0&-1&0
 \end{matrix}
\right).
\end{equation}
%%%%%%%%%%%

\subsubsection{Total spin $S=2$ of the Cooper pair} 
Here are five components now.
\begin{equation}
M_{2,2}
=
\sqrt{3} \left(
 \begin{matrix}
  0&0&1\\
  0&0&0\\
   0&0&0
 \end{matrix}
\right),
M_{2,1}
=
\sqrt{\frac{3}{2}} \left(
 \begin{matrix}
  0&-1&0\\
  0&0&1\\
   0&0&0
 \end{matrix}
\right),
\end{equation}

\begin{equation}
M_{2,0}
=
\frac{1}{\sqrt{2}} \left(
 \begin{matrix}
  1&0&0\\
  0&-2&0\\
   0&0&1
 \end{matrix}
\right),
M_{2,-1}=
\sqrt{\frac{3}{2}} \left(
 \begin{matrix}
  0&0&0\\
  1&0&0\\
   0&-1&0
 \end{matrix}
\right),
M_{2,-2}=
\sqrt{3} \left(
 \begin{matrix}
  0&0&0\\
  0&0&0\\
   1&0&0
 \end{matrix}
\right).
\end{equation}

It is convenient to write down explicitly: 
\begin{equation}
M_{2,2}\gamma =
\sqrt{3}
\left(
 \begin{matrix}
  1&0&0\\
  0&0&0\\
   0&0&0
 \end{matrix}
\right),
M_{2,1}\gamma =
\sqrt{\frac{3}{2}}
\left(
 \begin{matrix}
  0&1&0\\
  1&0&0\\
   0&0&0
 \end{matrix}
\right),
\end{equation}

\begin{equation}
M_{2,0}\gamma =
\frac{1}{\sqrt{2}} 
\left(
 \begin{matrix}
  0&0&1\\
  0&2&0\\
   1&0&0
 \end{matrix}
\right),
M_{2,-1}\gamma =
\sqrt{\frac{3}{2}}
\left(
 \begin{matrix}
  0&0&0\\
  0&0&1\\
   0&1&0
 \end{matrix}
\right),
M_{2,-2}\gamma =
\sqrt{3}
\left(
 \begin{matrix}
  0&0&0\\
  0&0&0\\
   0&0&1
 \end{matrix}
\right).
\end{equation}
%%%%%%%%%%%%

\subsection{Interaction}
The matrices introduced above will be used in the derivation of the fermion interaction term, describing the inter-valley Cooper pairing. The density-density interaction can be described by
\begin{eqnarray}\label{interaction_SM}
U = - \lambda \sum_{\alpha,\beta =-1}^{1}\int_{\bf k, k'} (\Psi_{1,\alpha, \mathbf{k}}^\dag  \Psi_{1,\alpha, \mathbf{k}'}) ( \Psi_{-1,\beta, -\mathbf{k}}^\dag\Psi_{-1, \beta, -\mathbf{k}'}),
\end{eqnarray}
%\begin{eqnarray}\label{interaction_1}
%U = - \lambda \sum_{\{\alpha\}}\int_{\bf k, k'} \Psi_{1,\alpha_1, \mathbf{k}}^\dag  \Psi_{-1,\alpha_2, -\mathbf{k}}^\dag\Psi_{-1, \alpha_3, -\mathbf{k}'} \Psi_{1,\alpha_4, \mathbf{k}'},~~~~
%\end{eqnarray}
where $ \lambda > 0$ is the interaction constant. Recall, that $\Psi_{s,\mathbf{k}} = [\Psi_{s, +1,\mathbf{k}}, \Psi_{s, 0,\mathbf{k}}, \Psi_{s, -1,\mathbf{k}}]^{T}$, where the first and second indices describe the valley and the band, respectively.
The above interaction term (\ref{interaction_SM}) can be conveniently rewritten as Ref.~[\ref{Nandkishore}]:
\begin{align}\label{SM_int}
U = - \frac{\lambda}{3}\sum_{S=0}^{2}\int_{\bf k, k'} [\Psi_{1,\mathbf{k}}^{\dag}  \mathbf{M}_S \gamma (\Psi_{-1, -\mathbf{k}}^\dag)^{\mathrm{T}}]
\cdot [\Psi^{\mathrm{T}}_{-1,-\mathbf{k}'}\gamma \mathbf{M}_S^{\dag} \Psi_{1,\mathbf{k}'}].
\end{align} 
As we will be investigating $S=0$ in what follows only, let us explicitly write down the corresponding contribution
\begin{eqnarray}\nonumber
U_0 &=& - \frac{\lambda}{3}\int_{\bf k, k'}[\Psi_{1,1,\mathbf{k}}^{\dag}\Psi_{-1,-1,-\mathbf{k}}^{\dag} - \Psi_{1,0,\mathbf{k}}^{\dag}\Psi_{-1,0,-\mathbf{k}}^{\dag} +\Psi_{1,-1,\mathbf{k}}^{\dag}\Psi_{-1,1,\mathbf{k}}^{\dag}  ]
\\
&\times& [\Psi_{-1,-1,-\mathbf{k}'} \Psi_{1,1,\mathbf{k}'} - \Psi_{-1,0,-\mathbf{k}'} \Psi_{1,0,\mathbf{k}'} + \Psi_{-1,1,\mathbf{k}'} \Psi_{1,-1,\mathbf{k}'} ].
\end{eqnarray} 
We emphasize that all bands contribute to the pairing channel. Taking both the Hamiltonian Eq. \ref{SM_nonint} and the interaction Eq. \ref{SM_int}, we obtain the Hamiltonian of the semimetal
\begin{equation}
\mathcal{H}_{\mathrm{tot}} = \int_{\mathbf{k}} [\Psi^{\dag}_{1,\mathbf{k}} v_{\mathrm{F}}\mathbf{S}\cdot \mathbf{k} \Psi_{1,\mathbf{k}} +
\Psi^{\dag}_{-1,\mathbf{k}} v_{\mathrm{F}}\mathbf{S}\cdot \mathbf{k} \Psi_{-1,\mathbf{k}} ] 
- \frac{\lambda}{3}\sum_{S=0}^{2}\int_{\bf k, k'} [\Psi_{1,\mathbf{k}}^{\dag}  \mathbf{M}_S \gamma (\Psi_{-1, -\mathbf{k}}^\dag)^{\mathrm{T}}]
\cdot [\Psi^{\mathrm{T}}_{-1,-\mathbf{k}'}\gamma \mathbf{M}_S^{\dag} \Psi_{1,\mathbf{k}'}].
\end{equation}
The gap-function is defined as
\begin{equation}
\boxed{
\mathbf{\Delta}_{S}^* =  \frac{\lambda}{3}\int_{\bf k} \langle \Psi_{1,\mathbf{k}}^{\dag}  \mathbf{M}_S \gamma (\Psi_{-1, -\mathbf{k}}^\dag)^{\mathrm{T}}\rangle
}
\end{equation}
where the components of $\mathbf{\Delta}_{S}$ are denoted by $\Delta_{S,m_S}$.
Define $\Delta_{00}\equiv \Delta$ for a scalar and $\mathbf{\Delta}_{1} = (\Delta_{1,1},\Delta_{1,0},\Delta_{1,-1})$ with $\mathbf{\Delta}_{2} = (\Delta_{2,2},\Delta_{2,1},\Delta_{2,0},\Delta_{2,-1},\Delta_{2,-2})$ for vector components.

Let us construct the BdG Hamiltonian 
\begin{equation}
\mathcal{H}_{\mathrm{BdG}} = \frac{1}{2} \int_{\mathbf{k}} \tilde{\Phi}^{\dag}_{\mathbf{k}}  \tilde{H}_{\mathrm{BdG}} (\mathbf{k}) \tilde{\Phi}_{\mathbf{k}} .
\end{equation}
The operators are given by
\begin{equation}
\tilde{\Phi}^{\dag}_{\mathbf{k}} = [\Psi^{\dag}_{1,\mathbf{k}}, \Psi^{\dag}_{-1,\mathbf{k}}, \Psi^{\mathrm{T}}_{1,-\mathbf{k}},  \Psi^{\mathrm{T}}_{-1,-\mathbf{k}} ]
\end{equation}
\\
and
\begin{eqnarray}
\tilde{H}_{\mathrm{BdG}} =
\left(
 \begin{matrix}
  v_{\mathrm{F}}\mathbf{S}\cdot \mathbf{k} -\mu S_0& 0 & 0 & \sum_{S=0}^2 \mathbf{\Delta}_S\cdot\mathbf{M}_S\gamma\\
  0& v_{\mathrm{F}}\mathbf{S}\cdot \mathbf{k} -\mu S_0& [\mathbf{\Delta}_1\cdot\mathbf{M}_1 -\sum_{S=0,2} \mathbf{\Delta}_S\cdot\mathbf{M}_S]\gamma &0\\
  0&\gamma [\mathbf{M}_1^{\dag}\cdot \mathbf{\Delta}^*_1 -\sum_{S=0,2}\mathbf{M}_S^{\dag}\cdot \mathbf{\Delta}_S^*] & v_{\mathrm{F}}\mathbf{S}^*\cdot \mathbf{k} +\mu S_0& 0\\
  \sum_{S=0}^2 \gamma \mathbf{M}_S^{\dag} \cdot\mathbf{\Delta}_S^* & 0 & 0 & v_{\mathrm{F}}\mathbf{S}^*\cdot \mathbf{k} +\mu S_0
 \end{matrix}
\right).~~~~~
\end{eqnarray}
We shall note the difference between $S=0,2$ and $S=1$ channels contributions to the $s$-wave inter-valley pairing. 
The symmetry of the $S=1$ channel is pseudospin-odd and only the intervalley-even is allowed. 
While the symmetry of the $S=0,2$ channels is pseudospin-even and the valley-odd pairing state is allowed.

It is convenient to perform a unitary transformation:
\begin{equation}
\tilde{\Phi}^{\dag}_{\mathbf{k}} = [\Psi^{\dag}_{1,\mathbf{k}}, \Psi^{\dag}_{-1,\mathbf{k}}, \Psi^{\mathrm{T}}_{1,-\mathbf{k}},  \Psi^{\mathrm{T}}_{-1,-\mathbf{k}} ] 
\rightarrow
\hat{\Phi}^{\dag}_{\mathbf{k}} = [\Psi^{\dag}_{1,\mathbf{k}}, \Psi^{\dag}_{-1,\mathbf{k}}, \Psi^{\mathrm{T}}_{1,-\mathbf{k}}\gamma,  \Psi^{\mathrm{T}}_{-1,-\mathbf{k}}\gamma ],
\end{equation}
which leads to
\begin{equation}
\mathcal{H}_{\mathrm{BdG}} = \frac{1}{2} \int_{\mathbf{k}} \hat{\Phi}^{\dag}_{\mathbf{k}}  \hat{H}_{\mathrm{BdG}}(\mathbf{k}) \hat{\Phi}_{\mathbf{k}}.
\end{equation}
Using $\gamma \mathbf{S}^*\gamma = -\mathbf{S}$, we now get
\begin{eqnarray}
\hat{H}_{\mathrm{BdG}} =
\left(
 \begin{matrix}
  v_{\mathrm{F}}\mathbf{S}\cdot \mathbf{k} -\mu S_0& 0 & 0 & -\sum_{S=0}^2 \mathbf{\Delta}_S\cdot\mathbf{M}_S\\
  0& v_{\mathrm{F}}\mathbf{S}\cdot \mathbf{k} -\mu S_0& -\mathbf{\Delta}_1\cdot\mathbf{M}_1 +\sum_{S=0,2} \mathbf{\Delta}_S\cdot\mathbf{M}_S & 0\\
  0 & -\mathbf{M}_1^{\dag}\cdot \mathbf{\Delta}^*_1 +\sum_{S=0,2}\mathbf{M}_S^{\dag}\cdot \mathbf{\Delta}_S^* & -v_{\mathrm{F}}\mathbf{S}\cdot \mathbf{k} +\mu S_0& 0\\
   -\sum_{S=0}^2 \mathbf{M}_S^{\dag} \cdot\mathbf{\Delta}_S^* & 0 & 0& -v_{\mathrm{F}}\mathbf{S}\cdot \mathbf{k} +\mu S_0
 \end{matrix}
\right).~~~~~
\end{eqnarray}
We emphasise, that the superconducting state is doubly degenerate. The inter-valley imbalance (analog of the Zeeman effect for conventional superconductors) will remove such degeneracy and suppress inter-valley 
pairing possibly via the small-momentum LOFF state. In the absence of such pair-breaking source, the BdG Hamiltonian splits into two $6\times 6$ blocks.

\section{BCS approach}
We will focus on the $S=0$ channel in what follows as it gives highest temperature for Cooper pairing Ref.~[\ref{Nandkishore}]. Here let us assume a spatial homogeneous order parameter. We will consider the spatial variation of the order parameter in the next section.
The system is described by the BdG Hamiltonian:
\begin{equation}
\mathcal{H}_{\mathrm{BdG}} = \int_{\mathbf{k}} \Phi^{\dag}_{\mathbf{k}}  H_{\mathrm{BdG}}(\mathbf{k}) \Phi_{\mathbf{k}},
\end{equation}
where the operators are now given by
\begin{equation}
\Phi^{\dag}_{\mathbf{k}} = [\Psi^{\dag}_{1,\mathbf{k}}, \Psi^{\mathrm{T}}_{-1,-\mathbf{k}}\gamma ]
\end{equation}
\\
and
\begin{equation}
H_{\mathrm{BdG}} =
\left(
 \begin{matrix}
  v_{\mathrm{F}}\mathbf{S}\cdot \mathbf{k} -\mu S_0& -\Delta S_0\\
   -\Delta^*S_0 & -v_{\mathrm{F}}\mathbf{S}\cdot \mathbf{k} +\mu S_0
 \end{matrix}
\right).
\end{equation}
recall $\Delta_{00} \equiv \Delta$. The self-consistency equation for this gap-function component reads
\begin{equation}
\frac{3\Delta}{\lambda} = T \sum_n\int_{\mathbf{k}}\left[ \frac{\Delta}{\omega_n^2+|\Delta|^2+\mu^2} + \sum_{s\pm 1}  \frac{\Delta}{\omega_n^2+|\Delta|^2+(v_{\mathrm{F}}k-s\mu)^2}\right].
\end{equation}
Again note that all bands contribute to the gap-function.
Summing up over the Matsubara frequencies
\begin{equation}
\sum_{n=-\infty}^{\infty}\frac{T}{[(2 n+1)\pi T]^2+a^2} = \frac{1}{2 a}\mathrm{tanh}\left(\frac{a}{2 T} \right)
\end{equation}
we obtain the self-consistency equation for the ordered state in the form

\begin{equation}\label{SQE_SM}
\boxed{
\frac{6}{\lambda} = \frac{K^3}{\sqrt{|\Delta|^2+\mu^2}} \mathrm{tanh}\left(\frac{\sqrt{|\Delta|^2+\mu^2}}{2 T} \right) +
\int_{\mathbf{k}} \sum_{s\pm 1} \frac{1}{\sqrt{|\Delta|^2+(v_{\mathrm{F}}k-s\mu)^2}} \mathrm{tanh}\left(\frac{\sqrt{|\Delta|^2+(v_{\mathrm{F}}k-s\mu)^2}}{2 T} \right)
}
\end{equation}

\subsubsection{Flat band}
Let us single out the flat-band contribution to the self-consistency equation:
\begin{equation}
\frac{6}{\lambda} = \frac{K^3}{\sqrt{|\Delta|^2+\mu^2}} \mathrm{tanh}\left(\frac{\sqrt{|\Delta|^2+\mu^2}}{2 T} \right).
\end{equation}
At $T=0$, one finds
\begin{equation}
|\Delta| = \mu\left\{ \frac{\lambda^2}{\lambda_c^2} - 1 \right\}^{1/2}, ~~~\lambda_c = \frac{6\mu}{K^3}.
\end{equation}
Solution exists provided the interaction constant is larger than the threshold $\lambda_c$. This is strong-coupling regime. 
Mathematically, at $\mu=0$ one obtains $|\Delta| = \lambda K^3/6$ at zero temperature.
On the other hand, at $\Delta \rightarrow 0$, one finds a solution for the transition temperature
\begin{equation}
T_p = \frac{\mu}{2 \mathrm{arcth}(\lambda_c/\lambda)},
\end{equation}
which only exists in the strong coupling regime $\lambda>\lambda_c$ as well. 

Note the momentum dependence of the flat-band contribution to the Green function (\ref{GF_1_SM})
\begin{align}
&\mathrm{Tr }\int_{\bf k} [S_0- (\mathbf{S}\mathbf{n}_{\bf k})^2][S_0- (\mathbf{S}\mathbf{n}_{{\bf k}-{\bf q}})^2]= 
\int_{\bf k}(\mathbf{n}_{\bf k} \mathbf{n}_{{\bf k}-{\bf q}})^2 = 
\left\{K^3 +\int_{\bf k} \frac{(\mathbf{n}_{{\bf k}}{\bf q})^2 - {\bf q}^2}{({\bf k}-\mathbf{q})^2} \right\} \approx K^3\left(1- 2 (6\pi^2)^{-2/3} \frac{{\bf q}^2}{K^2} \right).
\end{align}
estimated for $q \ll K$. It results in the spatial dependence of $\Delta$. We obtain flat-band defined Ginzburg-Landau equation at $\mu >|\Delta|$:
\begin{equation}\label{GL_equation_FB_SM}
\left(\frac{6}{\lambda}- \frac{K^3}{\mu} \mathrm{tanh}\frac{\mu}{2T} \right) \Delta - \alpha \frac{K}{\mu} \mathrm{tanh}\left(\frac{\mu}{2T}\right) \bnabla^2\Delta + \frac{3}{2\lambda_c}\frac{\mathrm{sinh}\frac{\mu}{T} -\frac{\mu}{T} }{\mathrm{cosh}^2\frac{\mu}{2T}} \frac{|\Delta|^2}{\mu^2} \Delta=0.
\end{equation}
\\
At $\mu>T$, we obtain
%\begin{equation}\label{GL_equation_FB_SM}
%\left(\frac{6}{\lambda}- \frac{K^3}{\mu}\right) \Delta - \alpha \frac{K}{\mu} \bnabla^2\Delta + \frac{K^3}{2\mu}\frac{|\Delta|^3}{\mu^2}=0,
%\end{equation}
%which can be slightly rewritten as 
\begin{equation}
\boxed{
\left( \frac{\lambda_c}{\lambda}-1 \right) \Delta - \frac{6\alpha}{K^2} \bnabla^2\Delta +\frac{|\Delta|^2 }{2\mu^2} \Delta =0
}
\end{equation}
The length
\begin{equation}
\xi = \frac{1}{K}\sqrt{\frac{\lambda}{\lambda-\lambda_c/ \mathrm{tanh}(\mu/2T) }},~~~ \lambda>\lambda_c
\end{equation}
defines characteristic length scale of variation of the order parameter. At $\mu\rightarrow T=0$, this length is of the order of Cooper pair size itself $K^{-1}$. 
We  can also obtain the supercurrent due to phase gradients
\begin{equation}
\mathbf{J} = 2\alpha \frac{e K}{\mu}|\Delta|^2 \bnabla \phi,
\end{equation}
where $e<0$ is the charge of electron.

\subsubsection{Flat and Dirac bands}
Let us now include Dirac band contribution given by the last term on the r.h.s. of Eq. \ref{SQE_SM} to the self-consistency equation.
Consider this contribution in limiting case $\mu=0$.  Evaluating at $T=0$, we find
\begin{equation}
\int_{0}^{\Lambda} \frac{dx}{2\pi^2 v_{\mathrm{F}}^3} \frac{2x^2}{\sqrt{|\Delta|^2+ x^2}} \mathrm{tanh}\left(\frac{\sqrt{|\Delta|^2+ x^2}}{2 T} \right)\bigg|_{T=0} 
\approx \frac{1}{2\pi^2 v_{\mathrm{F}}^3}\left\{ \Lambda^2 - |\Delta|^2 \ln \left|\frac{2\Lambda}{\Delta}\right|\right\},
\end{equation}
where $\Lambda$ is the ultraviolet cutoff for the Dirac band. Physically, it is responsible for the renormalization of the local flat-band contribution.
Taking both flat and Dirac band contributions at the charge neutrality point, the self-consistency equation reads
\begin{equation}
|\Delta| = \frac{\lambda K^3}{6} \left\{1- \frac{\lambda \Lambda^2}{12\pi^2v_{\mathrm{F}}^3} \left[1-\frac{|\Delta|^2}{\Lambda^2}\ln\left|\frac{2\Lambda}{\Delta}\right|\right] \right\}^{-1}.
\end{equation}
We will also assume that $\frac{\lambda \Lambda^2}{12\pi^2v_{\mathrm{F}}^3}<1$. It means that the 
Dirac band contribution itself does not tune the system to the strong interaction regime. We require at the charge neutrality 
$ \frac{\lambda \Lambda^2}{12\pi^2v_{\mathrm{F}}^3} \approx \frac{|\Delta| \Lambda^2}{(v_{\mathrm{F}} K)^3 } \ll 1,~ \left|\Delta/\Lambda\right|^2 \ll 1$.
We obtain
\begin{equation}
|\Delta| = \frac{\lambda K^3}{6} \frac{1 }{1- \frac{\lambda \Lambda^2}{12\pi^2v_{\mathrm{F}}^3}}.
\end{equation}
The interaction constant $\lambda$ is renormalized by the Dirac band contribution. 

Let us proceed to the limit of large chemical potential, $\mu \gg T>|\Delta|$.
Ignoring the Dirac band renormalization of the interaction constant by
subtracting the $\mu=0$ contribution from Eq. \ref{SQE_SM}, we find a self-consistency equation in the form
\begin{equation}\label{SQE_2_SM}
\frac{3}{\lambda}\left[1-\frac{\lambda}{\lambda_c}  \mathrm{tanh}\left(\frac{\mu}{2 T} \right) \right] - \nu_{\mathrm{nl}} \ln\left| \frac{2\mu}{T}\right|=0.
\end{equation}
Note the standard BCS logarithmic term due to Dirac band contribution. However, let's compare flat-band and Dirac band contributions in Eq. \ref{SQE_2_SM}:
\begin{equation}
\frac{\mathrm{tanh}\left(\mu/2 T\right) }{\lambda_c \nu_{\mathrm{nl}} \ln\left| 2\mu/T\right|} \sim \frac{1 }{\lambda_c \nu_{\mathrm{nl}}} \sim \left(\frac{K}{k_\mathrm{F}} \right)^3,
\end{equation}
where $k_\mathrm{F} = \mu/v_\mathrm{F}$ is the Fermi momentum. We find that the flat-band contribution determines the transition between metal and Cooper pairing states:
\begin{equation}
\boxed{
 \left(\frac{K}{k_\mathrm{F}} \right)^3 \gg 1
 }
\end{equation}

At $\lambda<\lambda_c$, the flat-band contribution does not lead to the pairing itself. In this weak coupling regime the Dirac band contribution gives rise to a solution
\begin{equation}
T_c = 2\mu \exp\left\{-\frac{3}{\lambda \nu_{\mathrm{nl}}}(1-\lambda/\lambda_c)\right\},~~~ \lambda<\lambda_c
\end{equation}
It is a BCS result for the transition temperature to the phase-coherent state. The transition is tuned by the long-range Dirac band contribution.

At $\lambda>\lambda_c$, the Dirac band contributes as a correction to the flat-band induced transition temperature $T_p$. We obtain the increase of temperature:
\begin{equation}
\tilde{T}_p = \frac{\mu}{2}\left\{\mathrm{arcth\left( \frac{\lambda_c}{\lambda} - \frac{\lambda_c \nu_{\mathrm{nl}}}{3}  \ln\left| \frac{2\mu}{T_p}\right| \right)} \right\}^{-1}
\approx T_p\left\{ 1 +  \frac{\lambda_c  \nu_{\mathrm{nl}}}{3}  \frac{2T_p}{\mu} \frac{ \ln\left| 2\mu/T_p\right|}{1-\lambda^2_c/\lambda^2} \right\}.
\end{equation}
However, we argue that $\tilde{T}_p$ is not a true 
transition to the phase-coherent state. %It is a temperature of transition between a metal and preformed Cooper pairing.

Let's include contribution to the spatial derivative of $\Delta$ originating from the Dirac band. Neglecting cubic terms in Eq. \ref{GL_equation_FB_SM}, we find usual BCS term as a correction to Eq. \ref{GL_equation_FB_SM} in the form
\begin{equation}
\left[ 1 - \frac{\lambda}{\lambda_c}  \mathrm{tanh}\left(\frac{\mu}{2 T} \right)  \left(1+ \frac{\alpha}{K^2} \bnabla^2 \right) - \frac{\lambda \nu_{\mathrm{nl}}}{3} 
 \left(\ln\left| \frac{2\mu}{T}\right| + \frac{7\zeta(3)}{48\pi^2}\frac{v_{\mathrm{F}}^2 }{T^2}\bnabla^2 \right) \right] \Delta   =0
\end{equation}
Let us compare gradient terms due to flat and Dirac bands:
\begin{equation}
\frac{\alpha}{K^2}\frac{\lambda}{\lambda_c}\mathrm{tanh}\left(\frac{\mu}{2 T} \right) \frac{3}{\lambda \nu_{\mathrm{nl}}} \frac{48\pi^2 T^2}{7\zeta(3)v_{\mathrm{F}}^2} \sim 
\frac{K}{k_\mathrm{F}}\frac{T^2 }{\mu^2}.
\end{equation}
We note that at $\mu\gg T$, the long-range Dirac band contribution to the gradient term dominates:
\begin{equation}
\frac{K}{k_\mathrm{F}}\frac{T^2 }{\mu^2}\ll 1
\end{equation}

However, can the long-range Dirac band contribution lead to the transition to phase-coherent state in the strong interaction case $\lambda>\lambda_c$?

%%%%%%%%%%%%%%%%%%%%%%
\section{Ginzburg-Landau formalism}
%%%%%%%%%%%%%%%%%%%%%%
We shall start with the partition function describing the system
\begin{equation}
Z = \int \mathcal{D}[\bar{\Psi},\Psi]\mathcal{D}[\bar{\Delta},\Delta] e^{-S[\bar{\Psi},\Psi; \bar{\Delta},\Delta]},
\end{equation}
with
\begin{equation}
S[\bar{\Psi},\Psi;\bar{\Delta},\Delta] =  -\int_{x} \bar{\Psi}(x) 
\left[
 \begin{matrix}
  (-\partial_{\tau} +\mu)S_0 - v_{\mathrm{F}}\mathbf{S}\cdot (-i\boldsymbol{\partial}_r) & \Delta(x)\\
   \bar{\Delta}(x) & (-\partial_{\tau} - \mu)S_0 + v_{\mathrm{F}}\mathbf{S}\cdot (-i\boldsymbol{\partial}_r) 
 \end{matrix}
\right]
\Psi(x)+ \frac{3}{\lambda}\int_{x} \Delta(x) \bar{\Delta}(x),
\end{equation}
where $x=(\tau, {\bf r})$. Integration over the fermionic variables gives $Z = \int \mathcal{D}[\bar{\Delta},\Delta] e^{-S[\bar{\Delta},\Delta]}$ in which now
\begin{equation}
\boxed{
S[\bar{\Delta},\Delta] =  -\int_x \ln\mathrm{det}\left[
 \begin{matrix}
  (-\partial_{\tau} +\mu)S_0 - v_{\mathrm{F}}\mathbf{S}\cdot (-i\boldsymbol{\partial}_r)& \Delta(x)\\
   \bar{\Delta}(x) & (-\partial_{\tau} - \mu)S_0 + v_{\mathrm{F}}\mathbf{S}\cdot (-i\boldsymbol{\partial}_r)
 \end{matrix}
\right]+ \frac{3}{\lambda}\int_{x} \Delta(x) \bar{\Delta}(x)
}
\end{equation}
Expansion over the bosonic field gives
\begin{eqnarray}\label{PF_main}
S[\bar{\Delta},\Delta] &=& S[0]+ \int_{x,x_1} \mathrm{Tr}\bigg\{\frac{1}{2}G(x-x_1) \underline{\Delta}(x_1)G(x_1-x)\underline{\Delta}(x) \\\nonumber
&+& \int_{x_2,x_3} \frac{1}{4}G(x-x_1)\underline{\Delta}(x_1)G(x_1-x_2)\underline{\Delta}(x_2) G(x_2-x_3)\underline{\Delta}(x_3)G(x_3-x)\underline{\Delta}(x)  \bigg\} +\frac{3}{\lambda}\int_{x} \Delta(x) \bar{\Delta}(x),~~~
\end{eqnarray}
where $G(x) = T\sum_{\omega}\int_{\bf k} G({\bf k}, i\omega_n) e^{i{\bf k \cdot r}-i\omega_n \tau}$ together with 
\begin{equation}
G({\bf k}, i\omega_n) = \left[
 \begin{matrix}
  (i\omega_n+\mu)S_0 - v_{\mathrm{F}}\mathbf{S}\cdot \mathbf{k} & 0 \\
   0 & (i\omega_n-\mu)S_0 + v_{\mathrm{F}}\mathbf{S}\cdot \mathbf{k}
 \end{matrix}
\right]^{-1}
\end{equation}
and
\begin{equation}
\underline{\Delta}=
\left(
 \begin{matrix}
0& \Delta(x)\\
   \bar{\Delta}(x) &0
    \end{matrix}
\right).
\end{equation}
To proceed, we shall first consider the flat-band contribution to the GL action and then analyze both flat and Dirac band contributions. 

%%%%%%%%%%%%%%%%%%%%%% BEGIN FIGURE %%%%%%%%%%%%%%%%%%%
\begin{figure}[t!]
\centering
\includegraphics[width=6.0cm]{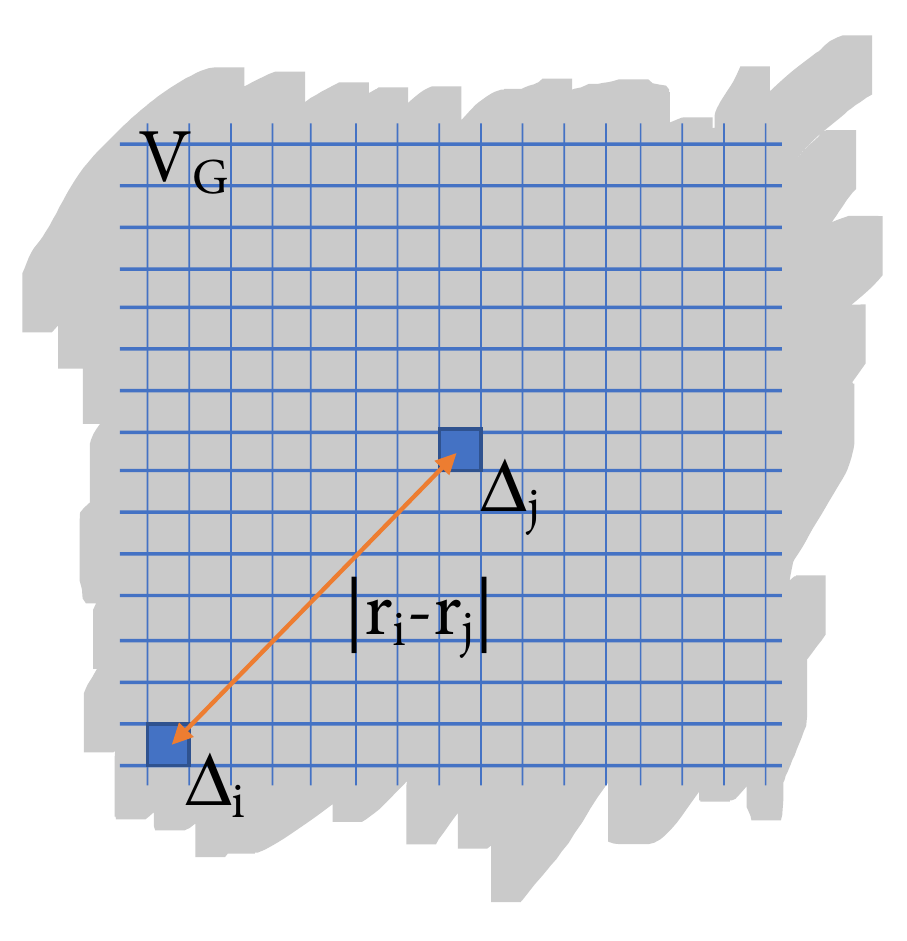}
\caption{\label{fig1_SM} The system is split into grains (segments) of volume $V_G$. The flat-band contribution supports Cooper pair formation on each grain. 
Hence the gap functions $\Delta_i$ with different phases (and generally, with different amplitudes as well) are defined on each grain. Delocalized contributions, which originate from the Dirac bands, provide long-range coupling between different grains.}
\end{figure}
%%%%%%%%%%%%%%%%%%%%%% END FIGURE %%%%%%%%%%%%%%%%%%%

\section{Mean field approach}
\subsection{Preformed pair state: Flat band only}

Consider flat-band contribution to the Cooper pairing of only. We start with expression \ref{PF_main}, in which we keep flat-band contribution
$
Z = \int \mathcal{D}[\Delta, \bar{\Delta}] e^{-S_{\mathrm{loc}}[\Delta, \bar{\Delta}]}
$ and using
\begin{equation}\label{GF_LOC_SM}
G_{\mathrm{loc}}(\mathbf{k},i\omega_n) = \frac{S_0- (\mathbf{S}\mathbf{n}_{k})^2}{i\omega_n +\mu}
\end{equation}
\\
We obtain
\begin{eqnarray}
S_{\mathrm{loc}}[\Delta, \bar{\Delta}] = \frac{V_{\mathrm{G}}}{T}\sum_i \left\{ \frac{3}{\lambda}|\Delta_i|^2 - \frac{T}{V_{\mathrm{G}}} \sum_n \ln \frac{\omega_n^2+\mu^2+|\Delta_i|^2}{\omega_n^2}\right\}
\end{eqnarray}
where $|\Delta_i|^2 \equiv \bar{\Delta}_i\Delta_i$. We will neglect derivatives of the order parameter. Using 
\begin{equation}
\prod_{n=-\infty}^{\infty} \left[ 1+ \frac{\mu^2+|\Delta_i|^2}{(2\pi T)^2(n+1/2)^2}\right] = \mathrm{cosh}^2\frac{\sqrt{\mu^2+|\Delta_i|^2}}{2T}
\end{equation}
we find
\begin{eqnarray}
S_{\mathrm{loc}}[\Delta, \bar{\Delta}] =  \frac{V_{\mathrm{G}}}{T}\sum_i \left\{ \frac{3}{\lambda}|\Delta_i|^2 - \frac{2T}{V_{\mathrm{G}}}  \ln \left|\mathrm{cosh} \frac{\sqrt{\mu^2+|\Delta_i|^2}}{2T} \right| \right\}
\end{eqnarray}

Extremum gives condition for the Cooper pair preformation
\begin{equation}
\frac{\lambda_c}{\lambda} \Delta_i  = \frac{\Delta_i \mu}{\sqrt{\mu^2+|\Delta_i|^2}} \mathrm{tanh}\frac{\sqrt{\mu^2+|\Delta_i|^2}}{2T},~~~~~~ \lambda_c = 6\mu V_{\mathrm{G}}
\end{equation}
In the limiting cases, we rederive
\begin{align}
&|\Delta_i| = \mu \sqrt{\frac{\lambda^2}{\lambda_c^2}-1},~~~ T=0\\
&T_p = \frac{\mu}{2 \mathrm{arcth}(\lambda_c/\lambda)},~~~ T\gg |\Delta_i|. 
\end{align}
where $T_p$ is the transition temperature between semimetal and preformed Cooper pair state. It is convenient to proceed in the limit $\mu\gg |\Delta_i|$. Expansion gives
\begin{eqnarray}
S_{\mathrm{loc}}[\Delta, \bar{\Delta}] -S_{\mathrm{loc}}[0] = \frac{V_{\mathrm{G}}}{T}\sum_i \left\{a |\Delta_i|^2 +\frac{b}{2} |\Delta_i|^4\right\}
\end{eqnarray}
where 
\begin{eqnarray}
a &=& 3\left( \frac{1}{\lambda} - \frac{1}{\lambda_c} \mathrm{tanh}\frac{\mu}{2T}\right),\\
b &=& \frac{3}{4\lambda_c}\frac{\mathrm{sinh}\frac{\mu}{T} -\frac{\mu}{T} }{\mu^2 \mathrm{cosh}^2\frac{\mu}{2T}}
\end{eqnarray}
We will be using the limiting case, $\mu> T$. We can estimate

\begin{align}
\boxed{
a = 3\left( \frac{1}{\lambda} - \frac{1}{\lambda_c}\right),~~~b = \frac{3}{2\mu^2\lambda_c}
}
\end{align}
We note that $\lambda = \lambda_c$ defines condition for the interaction strength for the transition to the preformed pair state. 
Solution for the preformed pairs $|\Delta_i| = \sqrt{-a/b} \propto \sqrt{\lambda - \lambda_c} $ exists provided
\begin{equation}
\lambda > \lambda_c
\end{equation}

\subsection{Preformed pair state: localized and delocalized contributions}
We start with
\begin{equation}
Z = \int \mathcal{D}\Delta \mathcal{D}\Delta^* e^{-S[\Delta, \Delta^*]},
\end{equation}
in which
\begin{eqnarray}
S[\Delta, \Delta^*]  &=& S_{\mathrm{loc}}[\Delta, \Delta^*] +S_{\mathrm{nl}}[\Delta, \Delta^*]  \\
&=&  \frac{V_{\mathrm{G}}}{T}\sum_i \left\{a |\Delta_i|^2 +\frac{b}{2} |\Delta_i|^4\right\} - \frac{\nu_{\mathrm{nl}}}{2v_{\mathrm{F}}}V^2_{\mathrm{G}}
T\sum_{n}\sum_{i\neq j}\frac{e^{-\frac{2|\omega_n|}{v} |\mathbf{r}_i - \mathbf{r}_j|}}{|\mathbf{r}_i - \mathbf{r}_j|^2}(\Delta_i \Delta_j^* + \Delta_i^* \Delta_j).
\end{eqnarray}
Summation over the Matsubara frequency gives
\begin{equation}
T\sum_{n} e^{-\frac{2|\omega_n| }{v_{\mathrm{F}}} |\mathbf{r}_i - \mathbf{r}_j|} =  \mathrm{sinh}^{-1}(2\pi T |\mathbf{r}_i - \mathbf{r}_j|/v_{\mathrm{F}}).
\end{equation}

Consider mean field approximation. We consider homogeneous case. No vortices. Which allows to consider the grain with the local parameter $\Delta_0$, which is coupled to the mean field order parameter.
\begin{eqnarray}
\sum_{i\neq j}\frac{\Delta_i \Delta_j^* + \Delta_i^* \Delta_j}{|\mathbf{r}_i - \mathbf{r}_j|^2 \mathrm{sinh}(\frac{2\pi T}{v_{\mathrm{F}}} |\mathbf{r}_i - \mathbf{r}_j|)} \rightarrow
2\sum_{i\neq 0}\frac{\langle\Delta_i\rangle \Delta_0^* + \langle \Delta_i^*\rangle \Delta_0}{|\mathbf{r}_i - \mathbf{r}_0|^2 \mathrm{sinh}(\frac{2\pi T}{v_{\mathrm{F}}} |\mathbf{r}_i - \mathbf{r}_0|) }+
\sum_{i\neq j} \sum_{j \neq 0} \frac{\langle\Delta_i\rangle \Delta_j^* + \langle \Delta_i^*\rangle \Delta_j}{|\mathbf{r}_i - \mathbf{r}_j|^2 \mathrm{sinh}(\frac{2\pi T}{v_{\mathrm{F}}} |\mathbf{r}_i - \mathbf{r}_j|) },
\end{eqnarray}
where the order parameter is self-consistently defined by
\begin{eqnarray}
\langle \Delta_0 \rangle = \frac{\int \mathcal{D}\Delta \mathcal{D}\Delta^*\Delta_0 e^{-S[\Delta, \Delta^*]} }{\int \mathcal{D}\Delta \mathcal{D}\Delta^* e^{-S[\Delta, \Delta^*]} }.
\end{eqnarray}

In the mean field approximation, we obtain integration over $\Delta_0$ and $\Delta_0^*$ only:
\begin{eqnarray}
\langle \Delta_0 \rangle = \frac{\int d \Delta_0 d \Delta_0^* \Delta_0 e^{-S_{\mathrm{MF}} [\Delta_0, \Delta_0^*]} }{\int d \Delta_0 d \Delta_0^* e^{-S_{\mathrm{MF}} [\Delta_0, \Delta_0^*]} },
\end{eqnarray}
where
\begin{equation}
S_{\mathrm{MF}}[\Delta, \Delta^*] =  \frac{V_{\mathrm{G}}}{T} \left\{a |\Delta_0|^2 +\frac{b}{2} |\Delta_0|^4\right\} - \frac{\nu_{\mathrm{nl}}}{v_{\mathrm{F}}}V^2_{\mathrm{G}}
\sum_{i\neq 0}\frac{\langle\Delta_i\rangle \Delta_0^* + \langle \Delta_i^*\rangle \Delta_0}{|\mathbf{r}_i - \mathbf{r}_0|^2 \mathrm{sinh}(\frac{2\pi T}{v_{\mathrm{F}}} |\mathbf{r}_i - \mathbf{r}_0|) }.
\end{equation}
Consider spatial homogeneous mean-field solution $\langle\Delta_i\rangle \rightarrow \langle\Delta_0\rangle \equiv \langle\Delta \rangle$.
Hence, we substitute summation with the integration over the coordinate as
\begin{equation}
V_{\mathrm{G}}\sum_{i\neq 0} \frac{1}{|\mathbf{r}_i - \mathbf{r}_0|^2 }\frac{1}{\mathrm{sinh}(\frac{2\pi T}{v_{\mathrm{F}}} |\mathbf{r}_i - \mathbf{r}_0|)} =\int
\frac{d\mathbf{r}}{r^2 \mathrm{sinh}(\frac{2\pi T}{v_{\mathrm{F}}} r) } =   \int_{\lambda_{\mathrm{F}}}^{\infty}
\frac{2\pi dr}{\mathrm{sinh}(\frac{2\pi T}{v_{\mathrm{F}}} r) } = -\frac{v_{\mathrm{F}}}{T}\ln \left|\mathrm{tanh}\frac{\pi T \lambda_{\mathrm{F}}}{v_{\mathrm{F}} }\right| \approx \frac{v_{\mathrm{F}}}{T} \ln \left| \frac{2\mu}{T}\right|,
\end{equation}
where $\lambda_{\mathrm{F}} = v_{\mathrm{F}}/2\pi \mu$ is the Fermi wave-length.
We obtain
\begin{equation}
S_{\mathrm{MF}}[\Delta, \Delta^*] =  \frac{V_{\mathrm{G}}}{T} \left\{a |\Delta|^2 +\frac{b}{2} |\Delta|^4 - \nu_{\mathrm{nl}}
[\langle\Delta\rangle \Delta^* + \langle \Delta^*\rangle \Delta] \ln \left| \frac{2\mu}{T}\right| \right\}
\end{equation}
and
\begin{eqnarray}
\langle \Delta \rangle = \nu_{\mathrm{nl}} \frac{V_{\mathrm{G}}}{T}  \ln \left| \frac{2\mu}{T}\right| \frac{\int d \Delta d \Delta^* \Delta
[\langle\Delta\rangle \Delta^* + \langle \Delta^*\rangle \Delta]
e^{-\frac{V_{\mathrm{G}}}{T} \left\{a |\Delta|^2 +\frac{b}{2} |\Delta|^4\right\} } 
}
{\int d \Delta d \Delta^* e^{-\frac{V_{\mathrm{G}}}{T} \left\{a |\Delta|^2 +\frac{b}{2} |\Delta|^4\right\} } }.
\end{eqnarray}
In cylindrical coordinates we obtain
\begin{eqnarray}
\langle \Delta \rangle = -\nu_{\mathrm{nl}} \langle\Delta\rangle \frac{V_{\mathrm{G}}}{T}  \ln \left| \frac{2\mu}{T}\right| \frac{\int d\rho \rho^2
e^{-\frac{V_{\mathrm{G}}}{T} \left\{a \rho^2 +\frac{b}{2} \rho^4\right\} } 
}
{\int d\rho \rho  e^{-\frac{V_{\mathrm{G}}}{T} \left\{a \rho^2 +\frac{b}{2} \rho^4\right\} } }.
\end{eqnarray}
For $\langle \Delta \rangle \neq 0$, we get

\begin{align}
&1 = \nu_{\mathrm{nl}} \frac{V_{\mathrm{G}}\langle |\Delta|^2\rangle }{T}  \ln \left| \frac{2\mu}{T}\right|, \\
&\langle |\Delta|^2\rangle \equiv \frac{\int_{0}^{\infty} dx x
e^{-\frac{V_{\mathrm{G}}}{T} \left\{a x +\frac{b}{2} x^2\right\} } 
}
{\int_{0}^{\infty} dx  e^{-\frac{V_{\mathrm{G}}}{T} \left\{a x +\frac{b}{2} x^2\right\} } }.
\end{align}
1. Consider weak coupling regime in which $a>0$. At the vicinity of the phase transition, we can neglect b-term. Using $\langle |\Delta|^2\rangle = T/aV_{\mathrm{G}}$, we obtain usual BCS result
\begin{equation}
\boxed{
T_c  = 2\mu e^{-\frac{3}{\lambda \nu_{\mathrm{nl}}}(1-\frac{\lambda}{\lambda_c})}, ~~~ \lambda<\lambda_c 
}
\end{equation}
\\
2. Consider strong coupling regime in which $a<0$. At $a^2V_{\mathrm{G}}/bT \gg 1$, using $\langle |\Delta|^2\rangle = -a/b$ we obtain our main result
\begin{equation}
T_c =  \frac{\lambda_c \nu_{\mathrm{nl}}}{3}\mu \left( 1- \frac{\lambda_c}{\lambda}\right)\ln \left| \frac{2\mu}{T_c}\right| ~~~ \lambda_c\gg \lambda-\lambda_c>0
\end{equation}
with the log-accuracy, we find
\begin{equation}
\boxed{
T_c =  \frac{\lambda_c \nu_{\mathrm{nl}}}{3}\mu \left( 1- \frac{\lambda_c}{\lambda}\right)\ln \left| \frac{6}{\lambda_c \nu_{\mathrm{nl}} \left( 1- \lambda_c/\lambda\right)}\right| 
}
\end{equation}

\subsection{General expression}
For numerical purposes, we can use general expression for the flat-band contribution to the functional.
\begin{align}
&1 = \nu_{\mathrm{nl}} \frac{V_{\mathrm{G}}\langle |\Delta|^2\rangle }{T}  \ln \left| \frac{2\mu}{T}\right|, \\
&\langle |\Delta|^2\rangle \equiv \mu^2 \frac{\int_{0}^{\infty} dx x
e^{-Y } }
{\int_{0}^{\infty} dx  e^{-Y }}, \\
& 
Y =  \frac{\lambda_c}{\lambda} \frac{\mu x}{2 T} -2  \ln\left[\mathrm{cosh}\left( \frac{\mu }{2T}\sqrt{1+x} \right)\right].
\end{align}
Hence, we simplify to

\begin{align}
\boxed{
1 = \frac{\lambda_c\nu_{\mathrm{nl}}}{3} \frac{\mu}{2T}  \ln \left| \frac{2\mu}{T}\right| \frac{\int_{0}^{\infty} dx x
e^{-Y } }
{\int_{0}^{\infty} dx  e^{-Y }},~~~~
Y =  \frac{\lambda_c}{\lambda} \frac{\mu }{2 T}x -2  \ln\left[\mathrm{cosh}\left( \frac{\mu }{2T}\sqrt{1+x} \right)\right]
}
\end{align}

\subsection{References in Supplemental Material}

\begin{enumerate}
\item \label{AGD_book_SM} A.A. Abrikosov, L.P. Gorkov, and I.E. Dzyaloshinskii, \textit{Methods of Quantum Field Theory in Statistical Physics} (Dover Publications, New York, 1975)

\item \label{Bradlyn_SM} B. Bradlyn, J. Cano et al., Science 353, 496 (2016). \textit{Beyond Dirac and Weyl fermions: Unconventional quasiparticles in conventional crystals}

\item \label{Nandkishore} Yu-Ping Lin and R. M. Nandkishore, Phys. Rev. B 97, 134521 (2018). \textit{Exotic superconductivity with enhanced energy scales in materials with three band crossings}

\end{enumerate}

\end{document}